\documentclass[12pt]{article}
\usepackage[utf8]{inputenc}
\usepackage{graphicx,psfrag,epsf}
\usepackage{booktabs}
\usepackage{textgreek}
\usepackage{threeparttable}
\usepackage{float}
\usepackage{amsmath,amsfonts,amsthm,bm} % Math packages
\usepackage{url}
\usepackage{transparent}
\usepackage{adjustbox}
\usepackage{flafter}
\usepackage{lscape}
\usepackage{caption}
\usepackage{subcaption}
\usepackage{graphicx}
\usepackage{geometry}
\usepackage{footnote} % For footnotes in a table
\makesavenoteenv{tabular} % For footnotes in table
\usepackage{verbatim}
\usepackage{natbib}
\usepackage{comment}
\usepackage{rotating}
\usepackage{hyperref}
\usepackage{mdframed}
\usepackage{lipsum}
\usepackage{array}
\newcolumntype{H}{>{\setbox0=\hbox\bgroup}c<{\egroup}@{}}
\usepackage{xcolor}
\usepackage{amsmath}
\usepackage{multirow}
\usepackage{bbm}
\usepackage{algorithmicx}
\usepackage{algorithm,algpseudocode}
\usepackage{longtable}

\newcommand{\blind}{0}

\begin{document}

\def\spacingset#1{\renewcommand{\baselinestretch}%
{#1}\small\normalsize} \spacingset{1}

%----------------------------------------------------------------------------------------
%	Title Page Section
%----------------------------------------------------------------------------------------
\def\spacingset#1{\renewcommand{\baselinestretch}%
{#1}\small\normalsize} \spacingset{1}

%%%%%%%%%%%%%%%%%%%%%%%%%%%%%%%%%%%%%%%%%%%%%%%%%%%%%%%%%%%%%%%%%%%%%%%%%%%%%%

\if0\blind
{
  \title{\bf Non-stationary Financial Risk Factors and Macroeconomic Vulnerability for the UK\thanks{
    The authors thank participants of the $10^{th}$ International Workshop on Applied Probability for their feedback. The usual disclaimer applies.}}
  \author{Katalin Varga\footnote{Corresponding author: vargaka@mnb.hu.} \hspace{.2cm}\\
    Central Bank of Hungary, Hungary \\
     \\
    Tibor Szendrei\\
    Department of Economics, Heriot-Watt University, UK.}
  \maketitle
} \fi

\if1\blind
{
  \bigskip
  \bigskip
  \bigskip
  \begin{center}
    {\LARGE\bf Title}
\end{center}
  \medskip
} \fi

%----------------------------------------------------------------------------------------
%	Abstract Section
%----------------------------------------------------------------------------------------
\begin{center}
    %\textbf{Early Draft! Please do not disseminate!}
\end{center}

\bigskip
\begin{abstract}
\noindent Tracking the build-up of financial vulnerabilities is a key component of financial stability policy. Due to the complexity of the financial system, this task is daunting, and there have been several proposals on how to manage this goal. One way to do this is by the creation of indices that act as a signal for the policy maker. While factor modelling in finance and economics has a rich history, most of the applications tend to focus on stationary factors. Nevertheless, financial stress (and in particular tail events) can exhibit a high degree of inertia. This paper advocates moving away from the stationary paradigm and instead proposes non-stationary factor models as measures of financial stress. Key advantage of a non-stationary factor model is that while some popular measures of financial stress describe the variance-covariance structure of the financial stress indicators, the new index can capture the tails of the distribution. To showcase this, we use the obtained factors as variables in a growth-at-risk exercise. This paper offers an overview of how to construct non-stationary dynamic factors of financial stress using the UK financial market as an example.
\end{abstract}

%----------------------------------------------------------------------------------------
%	Key words
%----------------------------------------------------------------------------------------

\noindent%
{\it Keywords:}  Systemic stress, Financial Stress Index, Dynamic Bayesian Factor Model, Non-stationary Factor Model, Financial System
Tail Risk. \\
\noindent
%{\it JEL:} 
\vfill

\spacingset{1.45} % DON'T change the spacing!

%----------------------------------------------------------------------------------------

\section{Introduction}

Monitoring and quantifying systemic stress in the financial system is a key exercise of macroprudential policy, since tracking the build-up of financial vulnerabilities is a basic component of financial stability. One way to do this is by the creation of indices that act as a signal for the policy maker. A common way to create such measures is by compressing available information into a few factors. While factor modelling in finance and economics has a rich history, most of the applications tend to focus on stationary factors. Nevertheless the seminal works \citet{escribano1994cointegration}, \citet{pena1987identifying} and \citet{pena2006nonstationary} offer us a framework to follow to model non-stationary factors as well. Due to the non-stationary dynamic factor models closeness to error correction modelling, non-stationary factor modelling has become more common in economics (see \citet{pena2004forecasting,poncela2021factor} and references therein). Surprisingly, non-stationary factor modelling had less impact on financial stress modelling. To the knowledge of the authors the only paper that attempts to utilise non-stationary factor models for financial stress modelling is \cite{szendrei2020fiss}. The motivation in their paper is that data transformations to make series stationary will change the information content of variables (echoing the findings of \citet{coulombe2021macroeconomic}). In this current paper we refine the method utilised in \citet{szendrei2020fiss} following the recent advancements in the literature \citep{poncela2021factor,castle2021modelling} and apply it to the UK market to create an index that encapsulates financial stress.

The starting point of using different types of factor models is the principal component analysis (PCA). The most important benefit of PCA is its ability to quantify the individual importance of a large number of indicators so that their weight is linked to their historical importance measured by the explained variance in the broader financial system. The PCA methodology has the advantage of being able to capture the interconnectedness of financial markets, which is a required feature of financial stress indices. This enables the user to interpret the importance of each indicator to the overall financial system. The more correlated a component is with the others, the higher the weight it is assigned in the model. The advantages of PCA modelling have seen it be utilised in economics and finance. Unfortunately, the baseline PCA methodology hinges on the data being stationary.

The key advantage of non-stationary factor modelling for financial stability, is that financial stress is characterised by a high degree of inertia. Modelling this with a stationary framework will only allow us to track sudden surges in financial stress, which is equivalent to tracking the start of a financial crisis. Furthermore, due to its stationary nature, such measures tend to revert to the mean very quickly unless one allows for heteroskedasticity.\footnote{Including heteroskedasticity in a dynamic factor model has its own myriad of problems such as the decision of what part of the model to have stochastic volatility.} Moving away from the stationary paradigm will tackle this without the need to include heteroskedasticity in our model. Furthermore, a non-stationary model allows for the tracking of gradual build-up of stress, which a stationary model would not necessarily be able to capture.

While the advantages of a non-stationary factor model for tracking financial vulnerabilities is clear, it is not trivial to construct such a measure. When building factor models and using principal component analysis (PCA) using non-stationary data, a distinction between theoretical and empirical covariance should be made.  Furthermore, a limit theorem is needed to relate them, which may not exist if the variables are non-stationary. This is on account of spurious correlation, which biases the weights of our model. In particular, the empirical mean and variance of non-stationary time series need not be constant\citep{johansen2012analysis}. However, in the case of persistent, n-order integrated data, there exists the so-called generalised empirical covariance, which can be related to the theoretical covariance via a limit theorem \citep{pena2004forecasting,pena2006nonstationary}.

This paper will offer an overview of how to construct non-stationary dynamic factors of financial stress. We will then use the non-stationary factor modelling framework to construct the UK Financial Stress Index (UKFSI) for the United Kingdom using several variables pertaining to the UK financial system. The choice of the UK economy for this exercise stems from several key factors. Firstly, the UK boasts robust and diverse financial markets, offering rich datasets conducive to thorough analysis. Secondly, the availability of multiple stress indices constructed specifically for the UK economy, such as the Country-Level Index of Financial Stress (CLIFS), and Sovereign Composite Indicator of Systemic Stress (SovCISS), facilitates a comprehensive comparison of methodologies and their respective performances.

In assessing the performance of our stress indices, we utilise a growth-at-risk metric, providing a comprehensive evaluation of their efficacy in capturing downside risk and potential vulnerabilities within the financial system. This method enables us to identify areas for enhancement and refinement, crucial for informing policy decisions effectively. Furthermore, our paper pioneers a comparative ``horse-race'' of various stress index methodologies through a growth-at-risk exercise, marking the first such endeavour in academic literature.

Our investigation uncovers intriguing insights into the efficacy of different stress indices across varying time horizons. While factor-based indices demonstrate notable effectiveness in short-term forecasting, the CISS emerges as particularly adept at longer-term forecasting, notably at a one-year horizon. This discrepancy underscores the necessity of considering diverse methodologies in stress index construction, as each approach unveils unique facets of risk. Specifically, the CISS methodology excels in identifying systemic risk, elucidating situations where multiple markets are collectively impacted, thus prolonging crises. Consequently, we advocate for perceiving the various stress index methodologies not as substitutes, but rather as complementary tools in comprehending the multifaceted nature of financial stress. This underscores the complexity inherent in measuring financial stress and underscores the imperative of adopting a nuanced, multifaceted approach to risk assessment in financial markets.

The paper is structured the following way: Section 2 gives a brief overview of financial stress and the difficulties pertaining to creating a comprehensive measure of it. This section also introduces the variables used in this paper to create a financial stress index for the UK economy. Section 3 describes the methodology of non-stationary factors. It gives an overview of how to capture non-stationary factors, incorporating the latest advancements in methodology. This section also describes how to select the optimal number of factors, before describing how non-stationary factor models are estimated. Section 4 describes the financial stress index for the UK and the optimal number of factors. This section also describes how we will evaluate the performance of the index using a growth-at-risk exercise. On account of growth-at-risk being done on GDP data, which is quarterly, we also create a monthly measure of GDP for the UK economy. Finally, we compare the performance of the stress index against other measures of financial stress created for the UK economy.

%\begin{itemize}
    %\item Paper reviews how risk factors are created in the literature
    %\item Through this review a gap in the literature is found and an index to fill this gap is created
    %\begin{itemize}
    %    \item Stationary factor model
    %    \item Persistent factor model
    %    \item Stationary CISS type model
        %\item Persistent CISS type model (NEW)
    %\end{itemize}
    %\item The paper offers a horse-race of the methods using a Growth-at-Risk excercise (To the knowledge of the authors this is the first paper that compares the different type of indices)
    %\item The UK economy is chosen to do this excercise: (1) Rich markets; (2) Several different measures have been constructed for the UK economy (UKFSI, CLIFS, SovCISS)  
%\end{itemize}

\section{Financial stress and its measures}
%Financial Stress
The aim of the UKFSI is to capture financial stress in the financial system of the United Kingdom. To identify financial stress, we first need to define what it is we are interested in measuring. Doing so will shine light on what are aspects of financial stress one is interested in measuring, which in turn guides variable selection. Throughout this text we refer to financial stress as the realised level of risk in the financial system. This materialisation of risk can be measured with a continuous variable, with extreme values occurring during crises events. In situations of extreme financial risk, there is a notable shift in investors' inclination towards holding less risky assets, known as flight to quality, or holding liquid assets that would allow them to adjust their position on short notice, which is known as flight to liquidity \citep{caballero2008collective,hakkio2009financial}. A comprehensive measure of financial stress should not only encompass variables reflecting flight to quality and liquidity but also include indicators that relate to why these shifts in investor preferences occur, i.e. variables that gauge rising uncertainty and information asymmetry.

Elevated uncertainty in financial markets stems from asset valuation ambiguity and the behaviour of other market participants. Information gaps worsen during financial stress, impacting borrowing costs and lending activity. Risk preferences are dynamic, with market participants tending to underestimate risks during bull markets and overestimate risks during high-stress periods \citep{aliber2015manias}.

Information asymmetry in financial markets occurs when one party possesses more information about a product than the other. Various scenarios can trigger information asymmetry, worsening during financial stress as doubts about the accuracy of information about other parties arise. The declining value of potential collateral during high financial stress exacerbates information gaps, contributing to higher borrowing costs and decreased lending activity \citep{gorton2009information}. Uncertainty about banks' solvency further diminishes lending activity, while on secondary markets, the widening information gap lowers the average asset price.

With these in mind, it is not surprising that during periods of heightened financial stress, investors shift to more liquid and better quality assets in an effort to limit potential losses. However shifting exposures to more liquid and less risky assets is rational at an individual level only. Simultaneous flocking to safer assets further exacerbates the problems on the market, potentially leading to more investors adjusting their portfolio. This contagion-like spread of financial stress can impact other financial markets as well leading to systemic stress \citep{hartmann2000systemic}.

Systemic stress poses a threat to the real economy by hindering financial intermediation. In times of elevated systemic stress, market participants face restricted options for hedging, potentially compelling them to confine their activities to the most liquid and least risky markets. This results in increased business costs for all firms, ultimately leading to a decline in investment. On account of this identifying financial stress is important.

Financial markets of each economy vary in terms of their development and depth, and as such different financial markets, and thus different variables, are likely to be important for different economies. To this end, we follow \citet{hollo2012ciss}, \citet{duprey2017dating}, and \citet{szendrei2020fiss}, by first specifying four financial markets that encapsulate the UK financial market. These are the government bond market, the corporate bond market, the foreign exchange market, and the stock market. The variables used to capture the different markets are presented in table (\ref{tab:varibs}). In this table we also show the p-values for the ADF test on the different variables. One can see that there is a mix of stationary and non-stationary variables.

The choice of candidate variables was limited by specific data criteria. Firstly, the UKFSI's goal is to measure financial stress promptly, hence only daily data was taken into account. Secondly, the indicators' fluctuations should reflect broad market trends. Lastly, the selected variables should effectively depict key aspects of financial stress. %\tibi{REWRITE THIS PARAGRAPH BECAUSE IT'S COPY PASTED UNTIL HERE}
The starting point was set before the great financial crisis of 2008, at January 1, 2005. This also influenced the variable choices.

\begin{table}[]
\centering
\caption{Variables for the Factor model and the P-values for the ADF test}
\label{tab:varibs}
\resizebox{\textwidth}{!}{
\begin{tabular}{l|l|c}
\hline
\multicolumn{1}{c|}{\textbf{Market}} & \multicolumn{1}{c|}{\textbf{Variable}} & \multicolumn{1}{c}{\textbf{ADF test}} \\ \hline \hline
\multirow{2}{*}{Govt. Bond Mkt.} & Risk premium on 10 year bond compared to US & 0.0094 \\
 %& Yield on 3 month T bill \\
 & Yield on 10 year government bond - Yield on 3 month & 0.1962 \\ 
 \hline
\multirow{5}{*}{FOREX} & EUR/GBP spot volatility ($\alpha=0.94$) & 0.0578 \\
 & USD/GBP spot volatility ($\alpha=0.94$) & 0.0388 \\
 & CHF/GBP spot volatility ($\alpha=0.94$) & 0.0059 \\
 & JPY/GBP spot volatility ($\alpha=0.94$) & 0.0520\\
 & Real Effective Exchange Rate volatility ($\alpha=0.94$) & 0.0404\\ 
 \hline
\multirow{6}{*}{Capital Mkt.} & CMAX of FTSE Small Cap (60 day window) & 0.0010\\
 & CMAX of FTSE 100 (60 day window) & 0.0010\\
 & CMAX of FTSE 350 (60 day window) & 0.0010\\
 & CMAX of FTSE 100 Euro (60 day window) & 0.0010\\
 & CMAX of FTSE 250 Euro (60 day window) & 0.0010\\
 & CMAX of FTSE 350 Euro (60 day window) & 0.0010\\ 
% & CMAX of FTSE World Mid Cap (60 day window) \\
% & CMAX of FTSE World Large Cap (60 day window) \\
 \hline
\multirow{5}{*}{Corporate Bond Mkt.} & S\&P UK Investment Grade Corporate Bond Index & 0.4343\\
 & S\&P UK 3-5 Years Investment Grade Corporate Bond Index & 0.1173\\
 & S\&P UK BBB Investment Grade Corporate Bond Index & 0.2770\\
 & S\&P UK A Investment Grade Corporate Bond Index & 0.3169\\
 & S\&P UK AA Investment Grade Corporate Bond Index & 0.1698\\
 \hline
\end{tabular}
}
\end{table}

%Equity Market
To capture the UK stock market six major stock market indices were included to capture the movement of the capital market in the UK. These indices can jointly capture flight to quality aspects as it is expected that FTSE small cap and FTSE 350 will react to financial stress faster than the FTSE 100, which includes larger firms in its composition. In essence, including various compositions of the stock market can help identify changes investors shifts in risk preferences.

Shifts in investor attitudes is difficult to ascertain from the raw series itself. Instead some measure of cumulative losses would contain the information relevant to measuring financial risk. To this end we opt to utilise the CMAX methodology, which measures cumulative losses of a series in a period of time, like in \citet{illing2006measuring}, \citet{hollo2012ciss}, or \citet{szendrei2020fiss}. The CMAX of a series is calculated as:

\begin{equation}
    CMAX_{t}=1-\frac{x_t}{\max [x\in(x_{t-j}|j=0,1,...,W)]}
\end{equation}

\noindent where $x_t$ is the stock market index at time $t$, and $W$ specifies the length of the window. In essence, the fraction looks at the current value of the stock market compared to its maximum value in the past $W$ days. The need to subtract this from 1, is so that we end up with an indicator that increases as cumulative losses increase. Note how the inclusion of multiple CMAX transformed stock indices allows us to measure increased uncertainty as cumulative losses are likely to ramp at at different rates in different composition of the index. The rolling window size was chosen 60 days in order to capture the most recent market developments. 

%Corporate bond market
It is important to note that firms can obtain financing through the bond market as well. As such we also include variables that can help identify the level of financial stress in the bond market. Nevertheless, just like in the case of stock market indices, it is rarely the level of corporate bond market indices that contains the relevant information pertaining to financial stress. Instead we follow \citet{chatterjee2022systemic} and look at the corporate bond spread relative to the relevant maturity,  5 and 10 year UK government bond respectively:

\begin{equation}
    CorpSpread_t=CorpIndex_t-GovtY_{t}
\end{equation}

Where $Y=5$ or $Y=10$ depending on the maturity of the corporate bond.  We calculate the bond spread for various corporate bond indices: S\&P UK Investment Grade Corporate Bond Index, S\&P UK 3-5 year Investment Grape Corporate Bond Index, S\&P UK BBB Investment Grade Corporate Bond Index, S\&P UK A Investment Grade Index, and finally the S\&P UK AA Investment Grade Corporate Bond Index. The reason for inclusion of various indices, is that investors flight to quality and liquidity is likely to manifest in the bond market for lower investment grades first. If the level of financial stress is high enough, more investors are likely to adjust their risk preferences, which in turn leads to the spread increase propagating in the market for higher investment grade bonds.

%Govt. Bond Market
The government bond market is captured by three variables: risk premium on the 10-year UK government bond compared to the US 10-year government bond, reference yield on the 3-month bond, and the reference yield on the 10-year bond. The risk premium measure can capture flight to quality episodes during times of country specific financial stress. 

The yield on the 3-month government bond and the 10-year government bond together represents the yield curve. It is known that during times of financial stress, the short-term outlook of an economy deteriorates as uncertainty increases. This in turn raises the short-term yield, potentially above the long-term yield, i.e. inverting the yield curve. As such, we include the difference of the two measures as it can help capture increased uncertainty in the market.

%FOREX
To capture stress related to the foreign exchange market we will use the spot market exchange rates of the GBP against the Euro, US dollar, Swiss Franc, and the Japanese Yen. We will also include the real effective exchange rate from BIS to account for other trade partners.

Just like in the case of stock and bond markets, the level of the exchange rate is rarely of concern when making statements of financial stress in the specific market. Instead, it is the volatility of the currency that is informative. Increased volatility in the foreign exchange market impedes trade as expected returns are more uncertain. Agents can alleviate this through the derivative market, or through insurance. Nevertheless, this inherently increases the cost of doing business and is one channel through which increased financial stress can impact the real economy.

To calculate the volatility an exponentially weighted standard deviation of the daily log change with a decay parameter of 0.94 was chosen. The historical standard deviation of daily log changes of currencies is widely accepted to measure exchange rate volatility due to its simplicity in calculation and the fact that it requires no further assumptions \citep{szendrei2020fiss}. Our aim is to capture financial stress in a timely manner hence exponential weighting was imposed on the standard deviation so that older observations have a lower impact on the current level of standard deviation. The EWSD was calculated with the following formula:

\begin{equation}
    EWSD(x)_t=\sqrt{\frac{\sum^T_{t=1}w_t(x_t-\Bar{x}^*)^2}{\sum^T_{t=1}w_t-\frac{\sum^T_{t=1}w_t^2}{\sum^T_{t=1}w_t}}}
\end{equation}

\noindent where $w_t$ is the weight and $\Bar{x}^*$ is the exponentially weighted moving average.

The decay parameter of 0.94 was chosen on account of it giving a series that is less likely to exhibit a tendency to 'rebound', i.e. after reaching a high value drop the following day, only to rise again. Such undesirable fluctuations prompted the selection of a delay factor of 0.94, aiming to mitigate erratic movements. On way to further mitigate this behaviour is to consider larger decay factors, but any larger values than 0.95 were deemed unsuitable as they would assign weights, albeit small, to observations from over 90 days ago. %An inherent drawback of this volatility measure is its backward-looking nature, lacking insight into investors' expectations. We hope to mitigate this by the inclusion of various measures, and jointly estimating all the markets.

%Market based motivation
Note that the variables are grouped by markets. This entails that one can estimate factors for the markets separately. To this end we will also create Market Factors, and compare their performance to the 'statistical' factors. The advantage of creating market specific factors is that the factors are guaranteed to relate to the respective markets. The disadvantage is that the estimated factors might be a suboptimal mix of the different risks in the financial markets. Furthermore, some markets (such as the foreign exchange market) are completely described by backwards looking measures, lacking insight into investors' expectations. This could lead to a market based factor that lags. Nevertheless, it is difficult to say ex-ante which approach is more suitable.

\section{Non-stationary factor methodology} \label{sec:DFM}
\subsection{Factor analysis of non-stationary variables}
Principal component analysis (PCA) is applicable for stationary or cointegrated variables \citep{johansen2012analysis}. For PCA in such a situation we use the maximum likelihood estimates (MLE) of the eigenvalues and eigenvectors to calculate the empirical covariance matrix of the data. If the empirical covariance matrix is the MLE of its population counterpart, then the eigenvalues and eigenvectors of the empirical covariance matrix can be used as estimates of the theoretical values. Note, that this relationship depends on the proper connection between the empirical and the population covariance (correlation) matrices. It is well-known since \citet{yule1926we} that this connection is missing in case of general I(1) variables: there is no relation between the theoretical and empirical concepts. Due to this, it is not possible to perform standard PCA in case of general integrated vectors, \textit{unless} its elements are I(1) and cointegrated. In the case of cointegrated variables, PCA can always be applied and a dynamic factor model, or common trend representation, is valid. This has been formalised in \citet{escribano1994cointegration} where the authors show that the components of the time series vector $X_t$ are cointegrated of order 0, with rank r if and only if: (1) $X_t$ can be transformed to a common trend representation; and (2) $X_t$ is driven by $n-r$ common factors that are jointly I(1) and $r$ factors that are jointly I(0). The later point is often described as the observed common factors representation. 

While informative, the above framework hinges on identifying cointegrating factors. In many cases when $(n-r)<<m$, it is difficult to know ex-ante how many cointegrating relationship there are in our data. As such, there is a need to generalise the above framework so as to allow for factor number selection as well. The key to this is the generalised covariance matrix \citep{pena2004forecasting,pena2006nonstationary}. In this section we give a brief overview of the method described in \citet{pena2006nonstationary}. Consider the following factor model: $X_t$ is an m-dimensional vector of observed time series, driven by a set of $r<m$ unobserved common factors.

\begin{equation} \label{eq:factor_model}
    X_t=Lf_t+\varepsilon_t
\end{equation}

Here $f_t$ is the r-dimensional vector of common factors, L is the $m \times r$ factor loading matrix, the sequence of noise $\varepsilon_t$ are normally distributed vectors, and have zero mean and a full rank diagonal covariance matrix ($\Sigma_\varepsilon$). Note, that in equation (\ref{eq:factor_model}) lagged values of $f_t$ may be present in which case L is a companion form matrix. With this extension, the diagonal structure of the noise covariance matrix is not a restrictive assumption. It also follows that all the common dynamics comes through the common factors, $f_t$. We assume that the common factors follow a vector autoregressive moving average, VARMA(p, q) model:

\begin{equation} \label{eq:VARMA}
    \Phi(B)f_t=d+\Theta(B)u_t
\end{equation}

\noindent where $\Phi(B)$ and $\Theta(B)$ are matrices of polynomials with size $r\times r$, B is the backshift operator, the roots of the determinantal equation $|\Phi(B)|=0$ can lie on or outside the unit circle, d is a $r \times 1$ vector of constants and $u_t$ is normally distributed, has zero mean and a full rank covariance matrix with no serial correlation. The components of the common factors $f_t$ can be either stationary or non-stationary, and the usual conditions for the invertibility of the VARMA models are assumed.

The factor model framework described here is equivalent to the Error Correction Model representation of \citet{engle1987co} and the VECM of \citet{johansen2012analysis}: the $n-r$ common factors capture the co-integrating relationship. The equivalence between the common factor representation and the ECM is key and underpins how one can create and interpret big-data cointegrating relationships. One downside of this relationship is that just like for the VECM, the choice of which variable to normalise for will have an impact on the results. Since the factor model is not identified under rotations, we assume the usual restriction on the loading matrix, $L'L=I$ \citep{aguilar2000bayesian}.

Assume that we also have stationary, zero mean factors beside the non-stationary one. Further, assume that there are $r_1$ common non-stationary factors $f_{1,t}$ and there are $r_2$ common stationary factors $f_{2,t}$. With this in mind we can formulate the key elements of equations (\ref{eq:factor_model}) and (\ref{eq:VARMA}) in a block structure:

\begin{equation} \nonumber
    \begin{split}
        f'_t&=(f_{1,t}',f_{2,t}') \\
        u_t'&=(u_{1,t}',u_{2,t}') \\
        L&=(L_1,L_2)\\
        \Sigma_u&=\begin{bmatrix}
        \Sigma_1 & 0\\
        0 & \Sigma_2
        \end{bmatrix}
    \end{split}
\end{equation}

While the $n-r$ common factors of, $f_{1,t}$, capture the co-integrating relationship, the remaining factors, $f_{2,t}$, capture the stationary common movements. $f_{2,t}$ contain important information for financial stress as well: they proxy volatility common across the variables. Together, the two groups of factors describe the distribution of financial stress with the $n-r$ factors capturing the "location" and the remainder factors capturing higher moments. It is important to note, that we are able to proxy the volatility implied by our risk factors without the need to introduce stochastic volatility in the model. It should be clear as well, that focusing solely on stationary factors misses the information in the first $n-r$ factors and can only capture volatility.

\subsection{Selecting the number of factors}

For all factor models, we have to determine the number of factors to estimate. In the presence of non-stationary data, this task is further complicated as we have to normalise the covariance matrix for said non-stationary variables. Following \citet{pena2006nonstationary}, we take the generalised covariance matrix of $X_t$, which is integrated of order d:

\begin{equation} \label{eq:GenCovMat}
    C_X(k)=\frac{1}{T^{2d+D}}\sum^T_{t=k+1}(X_{t-k}-\Bar{X})(X_t-\Bar{X})'
\end{equation}

\noindent where $\Bar{X}$ is the sample average and $D=\{0,1\}$ depending on the existence of a drift in $X_t$, and $k$ is the lag order. Just like the empirical covariance matrix is important for stationary data, this matrix will play a key role in non-stationary factor analysis.

%Assume that we also have stationary, zero mean factors beside the non-stationary one. Further, assume that there are $r_1$ common non-stationary factors $f_{1,t}$ and there are $r_2$ common stationary factors $f_{2,t}$. Suppose that the non-stationary factors, $r_1$, are generated as:

%\begin{equation*}
%    \begin{split}
%        (1-B)^df_{1,t}=c+v_t\\
%        v_t=\Psi(B)u_{1,t}
%    \end{split}
%\end{equation*}

%\noindent where d is a positive integer, denoting the order of integration, $c$ is a constant, $E(u_{1,t})=0$ and $Var(u_{1,t})=\Sigma_1>0$, $f_{1,-(d-1)}=f_{1,-(d-2)}=\cdots=f_{1,0}=0$, $\sum_i|| \Psi_i||<\infty$ and $rank(\psi_1)=r_1$ where $\Psi_1=\sum_{i=0}^\infty \Psi_i$.\footnote{Details on the limit theorem of the above factor model and the weak convergence of $C_X(k)$ can be found in \citet{pena2006nonstationary}.}

\begin{equation}
    \hat{M}_1(k)=\Big[ \sum^T_{t=k+1}X_tX_t' \Big]^{-1} \sum^T_{t=k+1} X_t X_{t-k}' \Big[ \sum^T_{t=k+1} X_{t-k}X'_{t-k} \Big]^{-1} \sum^T_{t=k+1} X_{t-k} X_t'
\end{equation}

Given the generalised covariance matrix in equation (\ref{eq:GenCovMat}) and the model in (\ref{eq:factor_model}), we can compute the squared sample canonical covariance matrices $\hat{M}_1(k)$. We use the fact that if we have r number of common factors as above, from which $r_1$ are non-stationary, m-r number of eigenvalues of $\hat{M}_1(k)$ will converge to zero in probability. Important to note that the order of integration of the non-stationary factors does not have to be the same. If the $r_1$ common non-stationary factors have different orders of integration, denoted by $d_i,$ where $i=1,...,r_1$, then we can use the following scaling matrix for $\hat{M}_1(k)$:

\begin{equation}
    \vartheta=diag\Big(\frac{1}{T^1},...,\frac{1}{T^{r_1}},...,\frac{1}{\sqrt{T}},...,\frac{1}{\sqrt{T}}\Big)
\end{equation}

Since the eigenvalues are continuous functions of the covariance matrix, the ordered eigenvalues $\hat{\lambda}_1\geq\cdots\geq\hat{\lambda}_m$ of $\hat{M}_1(k)$ can be used to compute a test statistic to obtain the number of factors.\footnote{Specifically the continuous mapping theorem can be used to prove the equivalence of the two eigenvalues. For more details see \citet{andersson1983distribution}.} Specifically, this test statistic $S_{m-r}$, which is asymptotically $\chi^2_{(m-r)^2}$ distributed, is calculated as:

\begin{equation}
    S_{m-r}=-(T-k)\sum_{j=1}^{m-r}log(1-\hat{\lambda}_j)
\end{equation}

Note that since $S_{m-r}$ depends on $M(k)$, the lag selected in $M(k)$ will have an influence on the critical value. To this end, the $S_{m-r}$ statistic is calculated for a variety of lags. The key advantage of using the $M(k)$, is that it can identify non-stationary and stationary factors jointly. 

The selection of the number of factors is not trivial and has a large influence on the estimated factors. To this end, having a procedure that reliably selects the optimal number of factors should not be understated. As such, compared to \citet{szendrei2020fiss}, utilising the factor number selection procedure provides additional robustness to the factor model method.

\subsection{Estimating the Model}

To estimate the model, we follow the procedure laid out in \citet{pena2006nonstationary} which will give us the number of stationary and non-stationary common factors and the initial estimate of the factor loading matrix.

As first step we test for the number of factors, so we build the matrix $\hat{M}_1(k)$ for $k=1,…,K$ and perform the chi-square test as described before to arrive for the number of factors, r. Next, we compute the generalised covariance matrices $C_X (k)$, estimate their eigenvalues and eigenvectors and sort them as is done in principal component analysis. An initial estimate of the factor loading matrix $\hat{L}^0$ could be the first r eigenvectors of $C_X (1)$, the initial estimate of the common factors is $\hat{f}^0=(\hat{L}^0)'X_t$. Finally, we test which one-dimensional elements of the vector of common factors are non-stationary.

The model is estimated with maximum likelihood using a Kalman Filter and the EM algorithm. Both Bayesian and non-Bayesian estimation can be carried out \citep{durbin2012time}. 
Dynamic factor models are structurally equivalent with state space models where factors can be treated as latent state variables. From this follows that algorithms for estimating state space models can also be used to estimate dynamic factor models. The authors opt to use Bayesian estimation since it has already proved to be a good choice generating smooth factors \citep{szendrei2020fiss}.

\section{UK Financial Risk Index (UKFSI)}
\subsection{Estimated Factors}
%%%%%%%%%%%%%%%%%%%%%%%%%%%%%%%%%%%%%%%%%%%%%%%%%%%%%%%%%%%%
%%%%%%%%%%%%%%%      FACTOR NUMBER      %%%%%%%%%%%%%%%%%%%%
%%%%%%%%%%%%%%%%%%%%%%%%%%%%%%%%%%%%%%%%%%%%%%%%%%%%%%%%%%%%
As it is described in section (\ref{sec:DFM}), to test the number of factors, we first need to build the canonical covariance matrices $\hat{M}_1(k)$ for $k=1,...,K$ and then perform a chi-squared test to estimate the number of factors, r. We follow \cite{pena2006nonstationary} in using lags 1 through 5. Doing so, the test reveals the presence of 5 to 6 factors.

We use the 18 input variables as shown in table (\ref{tab:varibs}). Based on the canonical covariance matrices and the generalised covariance matrices we have chosen models up to 5 factors. Both the generalised and the canonical covariance calculations suggest that the optimal number of factors is 5-6. For the sake of simplicity we opt to use 5 factors. An other reason for choosing 5 rather than 6 factors is the fact that we are covering 4 markets. To check the persistence of the factors we ran an ADF test on all 5 factors, which we report in table (\ref{tab:FactorPers}). From the table it is clear that all statistical factors are non-stationary, and as such all portray a high degree of persistence.%\kati{persistence of factors (ADF)}%In the performance evaluation we were testing models with sets of factors 1 to 5. 

\begin{table}[]
\centering
\caption{Explained Variance of factors}
\label{tab:ExplVar}
\begin{tabular}{ccc}
\hline
r & Expl. Variance & Cumulative \\ \hline \hline
1 & 0.568 & 0.568 \\
2 & 0.149 & 0.717 \\
3 & 0.095 & 0.811 \\
4 & 0.063 & 0.874 \\
5 & 0.046 & 0.920 \\ 
\hline
\end{tabular}
\end{table}

\begin{table}[]
\centering
\caption{Factor number test}
\label{tab:CritVal}
\begin{tabular}{c|cc|ccc}
\hline
& \multicolumn{2}{c|}{Crit. Values} & \multicolumn{3}{c}{$S_{m-r}$ test given $k$} \\
$r$ & $q_{0.05}$ & $q_{0.95}$ & $k=1$ & $k=2$ & $k=3$ \\ 
\hline \hline
0 & 61.261 & 103.010 & 345.182* & 324.672* & 306.423* \\
1 & 46.595 & 83.675 & 278.449* & 259.980* & 243.718* \\
2 & 33.930 & 66.339 & 213.897* & 198.607* & 185.526* \\
3 & 23.269 & 50.998 & 149.414* & 137.643* & 128.074* \\
4 & 14.611 & 37.652 & 86.546* & 79.304* & 73.986* \\
5 & 7.962 & 26.296 & 24.593 & 22.257 & 21.464 \\
6 & 3.325 & 16.919 & 2.983 & 2.971 & 2.964 \\
7 & 0.711 & 9.488 & 1.985 & 1.975 & 1.971 \\
8 & 0.004 & 3.841 & 0.988 & 0.981 & 0.979 \\
\hline
\end{tabular}
\end{table}

% \usepackage{graphicx}
\begin{table}[]
\centering
\caption{P-values for the ADF test on the different factors}
\label{tab:FactorPers}
%\resizebox{\textwidth}{!}{%
\begin{tabular}{l|ccccc}
 & Factor 1 & Factor 2 & Factor 3 & Factor 4 & Factor 5 \\ \hline
Market Factors & 0.3661 & 0.0034 & 0.001 & 0.306 &  \\
Statistical Factors & 0.3645 & 0.403 & 0.2489 & 0.1916 & 0.3121
\end{tabular}%
%}
\end{table}

\subsection{Evaluating performance of factors}
\subsubsection{Evaluation methods}
%%%%%%%%%%%%%%%%%%%%%%%%%%%%%%%%%%%%%%%%%%%%%%%%%%%%%%%%%%%
%%%%%%%%%%%%%%%%%%%%      QR      %%%%%%%%%%%%%%%%%%%%%%%%%
%%%%%%%%%%%%%%%%%%%%%%%%%%%%%%%%%%%%%%%%%%%%%%%%%%%%%%%%%%%

When it comes to evaluating the performance of stress indices, there is an inherent problem of what method to use. Ideally, one would use the methodology of \citet{kaminsky1999twin}, to determine the signalling potential of the variable, but this approach would require the knowledge of crises timings. Seeing, how the key reason to construct such variables is to help identify stress events, using this method is not possible. For this reason, some stress index papers (like in \citet{szendrei2020fiss} and \citet{hollo2012ciss}) opt for a narrative approach for evaluation, namely looking at the signals and describing what events occurred around the signal. While this approach is appealing to policymakers, it has difficulties distinguishing the signals from noise. To this end, a threshold VAR is often used as a more mathematically robust way to evaluate the performance of the stress indicator (see for example \citet{chatterjee2022systemic}). While this method is undeniably better than a simple narrative approach for evaluation, it is not a “model free” approach for evaluation. In essence, the performance of the stress index is tied to the number of regimes the researcher models. This makes index comparison difficult, since different stress indices might perform better with different regime numbers. Furthermore, the number of regimes can change across the sample. To alleviate the problems of the narrative approach and the threshold model approach, we propose using quantile regression as the framework to evaluate and compare the out-of-sample performance of the different risk indices. The key insight, is that the proposed risk indices fit nicely into the growth-at-risk (GaR) framework proposed by \citet{adrian2019vulnerable}. To this end we will fit densities conditional and use density evaluation metrics commonly used in the forecasting literature.

GaR has been popularised by \citet{adrian2019vulnerable}, who advocate modelling GDP growth with a value-at-risk framework. The consequence of this is that downside risk of GDP can be captured by the lower quantiles of the GDP growth density. The authors show that downside risk of GDP growth evolves with the state of the financial markets. Capturing these non-linearities helps in modelling GDP growth around crises episodes such as the global financial crisis of 2008. This entails that policy makers can glean how vulnerable the economy is to shocks with the help of a GaR model. Estimates for such a model can be obtained by estimating the following equation using quantile regression:

\begin{equation}
    y_{t+h}=x_t' \beta(\tau)+\varepsilon_{t+h}
\end{equation}

\noindent for $t=1,...,T-h$, where $h$ refers to the forecast horizon and $\tau\in(0,1)$ is the estimated quantile. $x_t$ includes a constant, a lag of GDP, and the risk index. Note that when $h=1$, the GaR is simply a QAR(1) model of \citet{koenker2006quantile} with financial variables as explanatory variables. The canonical GaR of \citet{adrian2019vulnerable} uses quarterly GDP growth in conjunction with the NFCI. For euro area applications the CISS has been used frequently \citep{figueres2020vulnerable}.

To obtain the $\beta(\tau)$ of the model we need to minimise the weighted absolute value of the residuals:

\begin{equation} \label{eq:QR}
\begin{split}
    \hat{\beta}(\tau)=\underset{\beta(\tau)}{argmax}\sum^{T-h}_{t=1}\big[&I(y_{t+h}\geq x_t'\beta(\tau))) |y_{t+h}-x_t'\beta(\tau)|\tau \\
    &+ I(y_{t+h}< x_t'\beta(\tau))) |y_{t+h}-x_t'\beta(\tau)|(1-\tau) \big]
\end{split}
\end{equation}

\noindent where $I(\cdot)$ is the indicator function. Using $\beta(\tau)$ from the above equation with the model specified before we will have the following conditional quantile:

\begin{equation}
    \hat{Q}_{y_{t+h}|x_t}(\tau|x_t)=x_t'\hat{\beta}(\tau)
\end{equation}

\citet{koenker1978regression} shows that this is a consistent linear estimate of the quantile ($\tau$) of $y_{t+h}$ conditional on $x_t$. As such as we change the quantile, we will get a forecast for the different quantiles of GDP. By estimating the above equation for a grid of quantiles we can construct densities and use density forecasting measures to evaluate the out-of-sample performance of the models with different risk indices. This allows for a principled way to evaluate the performance of the different risk indices.

To check the forecast performance of the quantile estimator with different measures of financial stress, the quantile weighted CRPS (qwCRPS) of \citet{gneiting2011comparing} is chosen as a scoring rule. To calculate this measure, we first take the Quantile Score (QS), which is the quantile weighted residual, which is the weighted reisudal for a given forecast observation ($\hat{y}_{t+h}$), with the quantile weight being the one in equation (\ref{eq:QR}). Using the QS, the qwCRPS is calculated as:

\begin{equation}
    qwCRPS_(t+h)=\int_0^1w_i QS_(t+h,\tau) d\tau
\end{equation}

\noindent where $w_i$ denotes a weighting scheme to evaluate specific parts of the forecast density. Through different weighting schemes we can evaluate differences at different part of the distribution. Since \citet{adrian2019vulnerable} has shown that financial conditions are more important for the lower tails, a natural way to evaluate the different risk indices is how they improve the out of sample density fit at the lower tails. 

Along with the qwCRPS we will also look at in-sample fit of the quantiles. To this end we present the Akaike- and Bayesian Information Criteria as shown in \citet{jiang2014interquantile}. We will compare the performance of the factor models to the CISS (UK specific CISS, denoted as SovCISS) and the CLIFS.

%%%%%%%%%%%%%%%%%%%%%%%%%%%%%%%%%%%%%%%%%%%%%%%%%%%%%%%%%%%%
%%%%%%%%%%%%%%%%      MONTHLY GDP      %%%%%%%%%%%%%%%%%%%%%
%%%%%%%%%%%%%%%%%%%%%%%%%%%%%%%%%%%%%%%%%%%%%%%%%%%%%%%%%%%%

While the GaR is a powerful framework to assess tail risks related to the macroeconomy, it utilises GDP as a measure, which is inherently a quarterly measure. In contrast, risk indices are often much higher frequency (daily or weekly). We can follow \citet{adrian2019vulnerable}, \citet{figueres2020vulnerable}, and \citet{szendrei2023revisiting} and take quarterly aggregates, but this would throw away too much timely information. Another approach would be to utilise a Mixed-Frequency framework as done in \citet{ferrara2022high} or \citet{xu2023mixed}, however the different risk measures might end up having different polynomial structures which would make comparison of the performance more difficult. In particular, it would be difficult to ascertain where difference in performance stem from: improvements of one stress index might be on account of the polynomial structure estimation being better for one measure rather than the risk index being more informative. To this end we propose the following compromise: run a GaR with the different risk indices on monthly GDP values. While aggregating the indices is still required, monthly aggregates undoubtedly have less loss of information than quarterly ones.

Recently much research was conducted on mixed-frequency estimates of GDP, see \citet{koop2021nowcasting}, \citet{huber2023nowcasting}, \citet{koop2023reconciled}, and \citet{schorfheide2021real} among others. The main reason for this was the extremely volatile data of the COVID period and thus a need for timely estimates for GDP. We can utilise the inroads of this research for our evaluation purposes. To this end we will use \citet{koop2023reconciled} for calculating the monthly estimates of UK GDP.\footnote{The main reason of the model choice was that \citet{koop2023reconciled} also provides historical monthly estimates of GDP compared to models like \citet{schorfheide2021real}, which are providing monthly frequency nowcasts and forecasts solely.}

To obtain monthly GDP figures the authors start from quarterly GDP values based on expenditure and income approach and considering them as noisy observations of true GDP. \citet{koop2023reconciled} use noise restriction based on this assumption, so an assumption is made that the variance of true GDP is less than the variance of its noisy observation. Since UK has no income-based constant prices GDP readily available, we opt to use the production- and expenditure-based constant price GDP. We also adjust the variance ratio priors based on the data to accommodate this change.\footnote{The underlying dataset and details of the priors can be found in the Technical Appendix.} Our quarterly frequency equations based on this modification are the following:

\begin{equation}
\begin{split}
    \begin{bmatrix}
        GDP_{P,t}\\GDP_{E,t}
    \end{bmatrix}&=1_{2 \times 1} GDP_t+
    \begin{bmatrix}
        \varepsilon_{P,t} \\ \varepsilon_{E,t}
    \end{bmatrix} \\
    GDP_t&=\rho GDP_{t-1}+\varepsilon_{G,t}
\end{split}
\end{equation}

\noindent where $GDP_{P,t}$ is production-based GDP, $GDP_{E,t}$ is expenditure-based GDP, and $GDP_{t}$ is true GDP. Following \citet{koop2023reconciled} we utilise the following reparametrisation:

\begin{equation}
    \xi_i=\frac{var(GDP)}{var(GDP_i)}
\end{equation}

This reparametrisation reflects the error in measurement hypothesis, namely that production- and expenditure-based GDP is equal to true GDP plus measurement error. This reparametrisation also enables us to set priors as intervals for the variance ratio $\xi_i$ as it is described in \citet{koop2023reconciled},
\noindent where $i\in\{P,E\}$, for the production- and expenditure based GDP. We opt to use the production-based GDP for the growth-at-risk estimation, since it does not contain taxes. The reason we use this GDP measure is that taxes are not closely linked to the performance of the economy. We assume that $0.35<\xi_P$ , $\xi_E<1.15$ based on the empirical variance of production- and expenditure-based GDP. Our prior interval for parameter of $\xi_P$ and $\xi_E$ is slightly larger compared to \citet{koop2023reconciled}.

The mixed frequency model can be summarised as:

\begin{equation}
\begin{split}
    y_t&=(X'_t,U_t,GDP_t,GDP_{P,t},GDP_{E,t}) \\
    y_t^Q&=\Delta_3\ln (Y_t) 
\end{split}
\end{equation}

\noindent where $Y_t^Q$ is the quarterly variable observed every third month, $U_t$ is the unemployment rate, that depends on $GDP$ but not on $GDP_P$ or $GDP_E$, and $X_t'$ is a set of monthly explanatory variables. We use the monthly explanatory variables of \citet{schorfheide2015real} and \citet{koop2023reconciled}: retail sales, inflation, industrial production, base rate, short-term interest rates, long-term interest rates, and stock prices. Exact definitions and data transformations are given in the appendix.

\begin{figure}
    \centering
    \includegraphics[width=\textwidth]{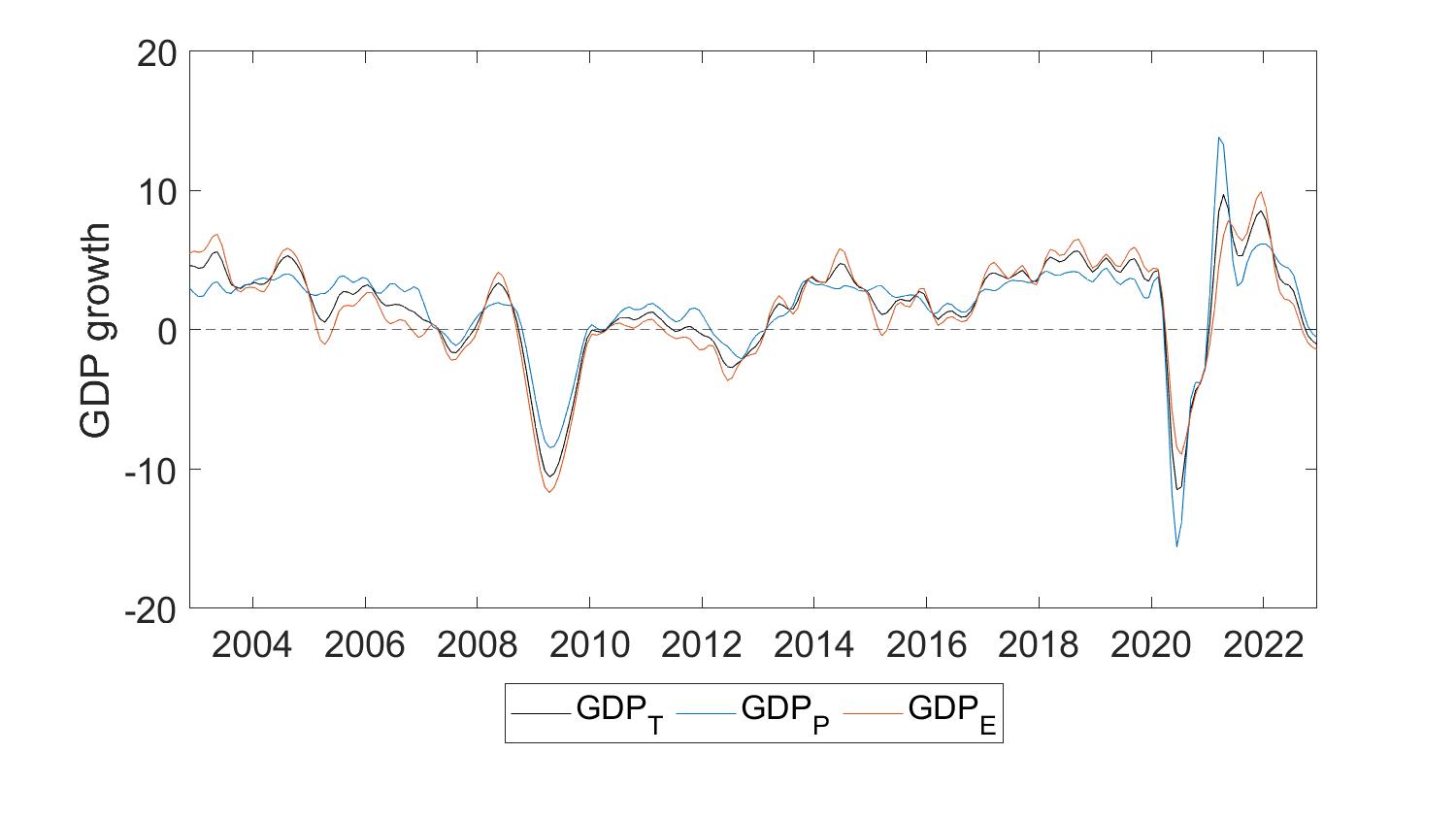}
    \caption{The different measures of monthly GDP}
    \label{fig:GDP_m}
\end{figure}

The estimated monthly GDP series are presented in figure (\ref{fig:GDP_m}). On can see from the figure that the general movement of all the monthly GDP measures follows an intuitive path: (1) During the 2008 crisis all measures dig below 0; (2) During the lockdown of COVID-19 pandemic, they portray strong negative growth, followed by; (3) a reversion in GDP when lockdown were lifted. We also note that $GDP_T$ is always between $GDP_E$ and $GDP_P$ as expected. 
\subsubsection{Evaluation Results}
%%%%%%%%%%%%%%%%%%%%%%%%%%%%%%%%%%%%%%%%%%%%%%%%%%%%%%%%%%%%
%%%%%%%%%%%%%%%%      EVAL RESULT      %%%%%%%%%%%%%%%%%%%%%
%%%%%%%%%%%%%%%%%%%%%%%%%%%%%%%%%%%%%%%%%%%%%%%%%%%%%%%%%%%%

\begin{figure}
    \centering
    \begin{subfigure}[b]{0.49\textwidth}
    \includegraphics[width=1.1\textwidth]{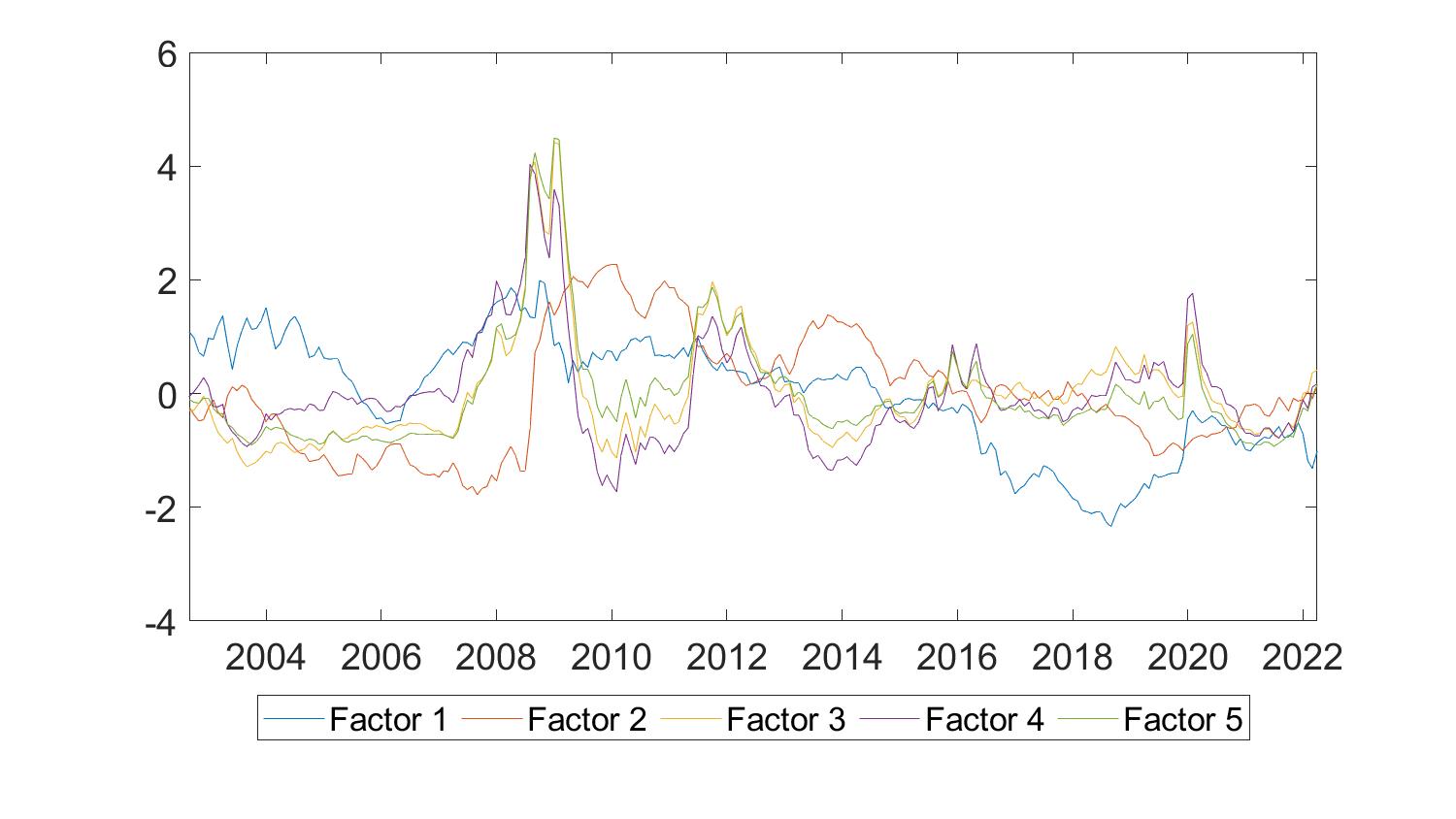}
    \caption{5 (statistical) factors}
    \label{fig:5fact}
    \end{subfigure}
    \begin{subfigure}[b]{0.49\textwidth}
    \includegraphics[width=1.1\textwidth]{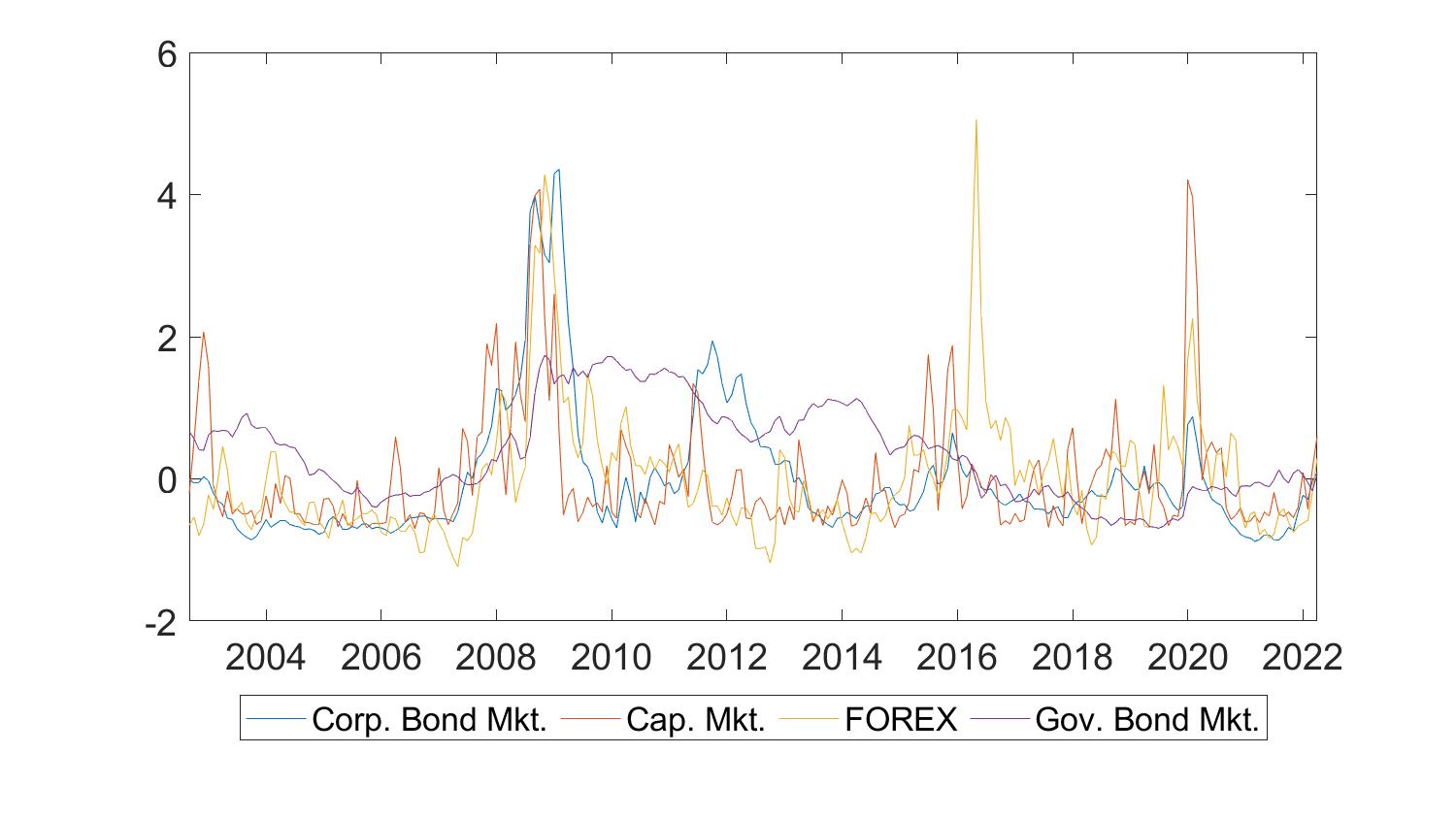}
    \caption{Market factors}
    \label{fig:mktfact}
    \end{subfigure}
    \caption{Individual factors}
\end{figure}

\begin{figure}
    \centering
    \includegraphics[width=\textwidth]{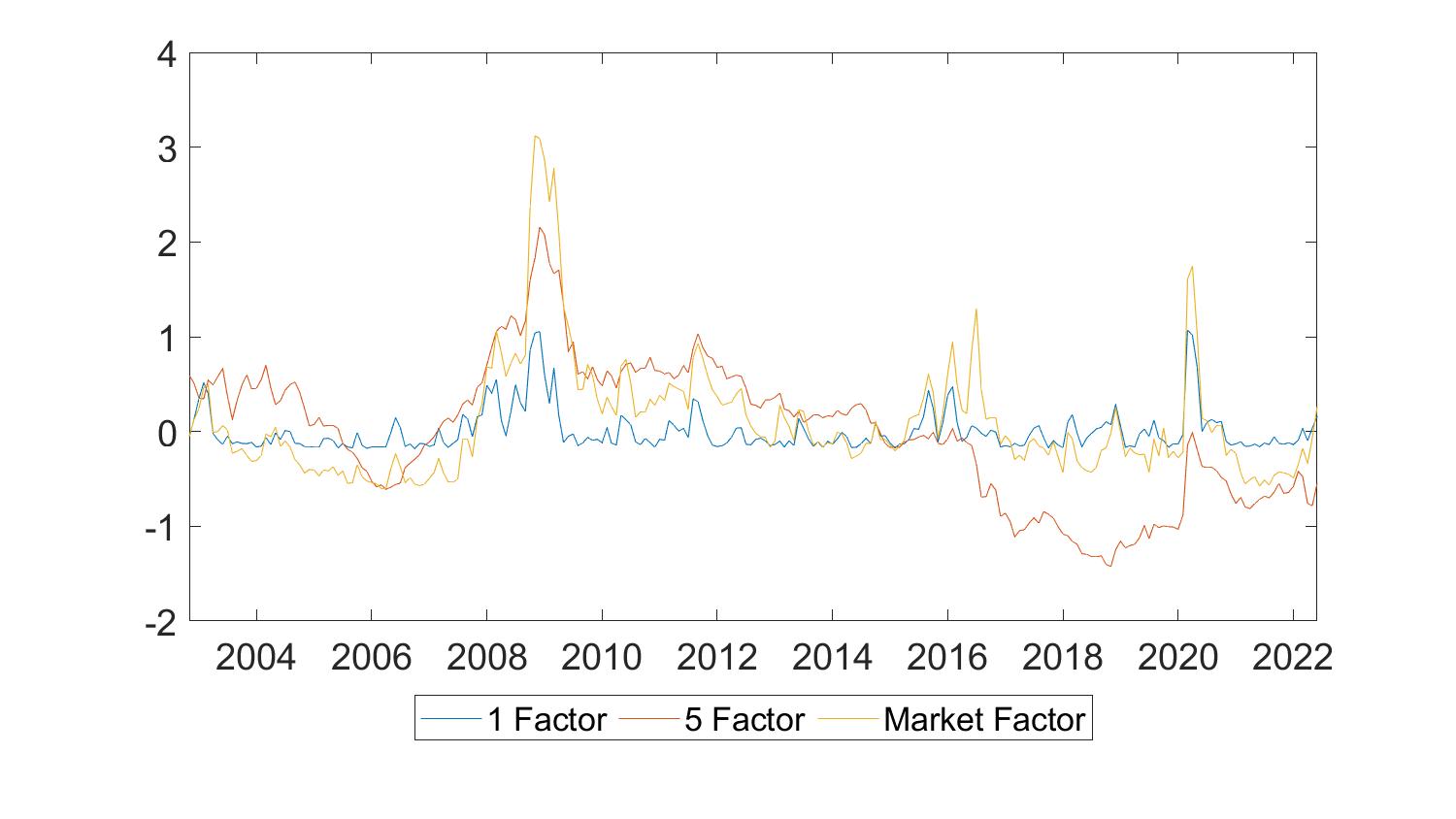}
    \caption{Combined factors}
    \label{fig:combfact}
\end{figure}

Before diving into the evaluation measures, we will present the estimated monthly factors. The statistical factors with 5 factors is shown in figure (\ref{fig:5fact}), and the market factors are shown in figure (\ref{fig:mktfact}). The combined versions of these factor models (as well as a 1 factor model) is shown in figure (\ref{fig:combfact}). These combined factors are currently calculating using a simple arithmetic mean. While it is possible to use other weighting schemes, such as exponential weighting (or BEKK GARCH) as done in \citet{chatterjee2022systemic}, doing so would make it difficult to ascertain where potential improvements stem from: aggregation method or factor methodology. To this end we feel that using the average is a sufficient way to aggregate the factors for our purpose of evaluation. Optimal method of aggregation remains an avenue for future research.

Comparing the ADF test results of the variables (shown in table (\ref{tab:varibs})) with the ADF test results of the factors (shown in table (\ref{tab:FactorPers}), we can see that the statistical factors are more persistent than the variables they are constructed from. Furthermore, the statistical factors are more persistent than the market factors (i.e. factor model created on each market individually). This results in smoother factors as shown in figures (\ref{fig:5fact}) and (\ref{fig:mktfact}). Given that policy makers would not like to minimise the chance of false positives, this smooth factor is a considerable advantage for the stress index constructed as statistical non-stationary factors. 

Looking at the different factors one can see that the proposed methodology and the market factors identify noticeably different factors. Although the scales of the identified factors are similar, we can see that the market factors are characterised by sudden spikes in its progression, while the statistical factors are smoother. This highlights that the statistical factor modelling methodology does not simply identify a factor for each market, but also grabs common information across the markets. This is particularly useful for measuring financial stress, as increased uncertainty is not necessarily contained to one particular market. On account of these, the statistical factors are expected to perform better when it comes to forecasting downside risk of GDP growth.

When it comes to evaluating the performance of the different indexes we will focus on 1 month ahead, 1 quarter ahead, half a year ahead, and 1 year ahead forecast. Table (\ref{tab:RAWresult}) shows the results of the different factors. Several points emerge from this table. First, the models with CLIFS and Market Factors is never chosen as the optimal model. The CLIFS doing worse is likely on account of the method using less financial variables than the CISS or the factor model. To this end, these results show that additional gains in modelling financial stress can be gained with with more information. The market factors not being selected showcases that it is better to model factors jointly, rather than separately for each market. This is largely driven by the fact that the reasons underlying financial stress are likely to occur in multiple markets jointly, and as such factors modelled on all financial variables has a better chance of uncovering such co-movements. 

Second, the model with the factors, provide better fit at shorter horizons than the model with CLIFS and SovCISS. However, the same cannot be said for longer horizons, where the SovCISS provides the best in sample and out of sample performance. This highlights that although the data these measures are constructed from are similar, the method of aggregation reveals different aspects of risk. In particular the methodology of the SovCISS helps reveal cases of systemic risk, i.e. situations where different markets are jointly impeded. Such crises have a longer lasting impact which explains power to forecast downside risk at longer horizons. As such, the different stress index methodologies are not substitutes but should instead be looked as complements. This also highlights that it is difficult to have an overarching measure of financial stress, stemming from the fact that different types of financial stress exists.

Third, of the factor models chosen, the 5 factor model yields the best performance especially at 3 and 6 month ahead horizon. Furthermore, when it comes to in-sample fit the 5 factor model is the best performing one with both AIC and BIC. Interestingly, the 1 factor model produces better out of sample results at the 1 month ahead horizon, nevertheless, the gains are marginal.

\begin{table}[]
\centering
\caption{In- and out-of-sample performance of the GaR with different risk measures}
\label{tab:RAWresult}
\begin{tabular}{llccccc}
\hline
 &  & 1 Factor & 5 Factors & Market Factors & CLIFS & SovCISS \\ \hline
\multicolumn{2}{l}{h=1} &  &  &  &  &  \\
 & AIC & 76.001 & \textbf{75.227} & 75.616 & 76.857 & 76.647 \\
 & BIC & 76.280 & \textbf{76.064} & 76.313 & 77.136 & 76.926 \\
 & $w_{centre}$ & \textbf{0.054} & 0.056 & 0.056 & 0.055 & 0.055 \\
 & $w_{left}$ & \textbf{0.085} & 0.087 & 0.088 & 0.089 & 0.088 \\
 %& $w_{right}$ & \textbf{0.139} & 0.147 & 0.145 & 0.141 & 0.141 \\ 
 \hline
\multicolumn{2}{l}{h=3} &  &  &  &  &  \\
 & AIC & 92.307 & \textbf{90.488} & 91.242 & 93.481 & 92.841 \\
 & BIC & 92.587 & \textbf{91.329} & 91.943 & 93.762 & 93.121 \\
 & $w_{centre}$ & 0.132 & \textbf{0.131} & 0.134 & 0.136 & 0.132 \\
 & $w_{left}$ & \textbf{0.210} & \textbf{0.210} & 0.218 & 0.230 & 0.222 \\
 %& $w_{right}$ & \textbf{0.333} & 0.338 & 0.336 & 0.339 & \textbf{0.333} \\ 
 \hline
\multicolumn{2}{l}{h=6} &  &  &  &  &  \\
 & AIC & 99.394 & \textbf{96.114} & 97.299 & 99.854 & 98.253 \\
 & BIC & 99.677 & \textbf{96.963} & 98.007 & 100.137 & 98.536 \\
 & $w_{centre}$ & 0.189 & \textbf{0.178} & 0.189 & 0.198 & 0.186 \\
 & $w_{left}$ & 0.324 & \textbf{0.297} & 0.324 & 0.340 & 0.310 \\
 %& $w_{right}$ & 0.466 & \textbf{0.451} & 0.461 & 0.484 & 0.464 \\ 
 \hline
  \multicolumn{2}{l}{h=12} &  &  &  &  &  \\
 & AIC & 101.900 & 98.965 & 99.897 & 101.254 & \textbf{97.207} \\
 & BIC & 102.188 & 99.830 & 100.618 & 101.542 & \textbf{97.496} \\
 & $w_{centre}$ & 0.237 & 0.221 & 0.233 & 0.231 & \textbf{0.184} \\
 & $w_{left}$ & 0.399 & 0.381 & 0.397 & 0.385 & \textbf{0.306} \\
 %& $w_{right}$ & 0.577 & 0.540 & 0.560 & 0.563 & \textbf{0.447} \\ 
 \hline
\end{tabular}
\end{table}

\begin{figure}
    \centering
    \begin{subfigure}[b]{0.49\textwidth}
        \includegraphics[width=\textwidth]{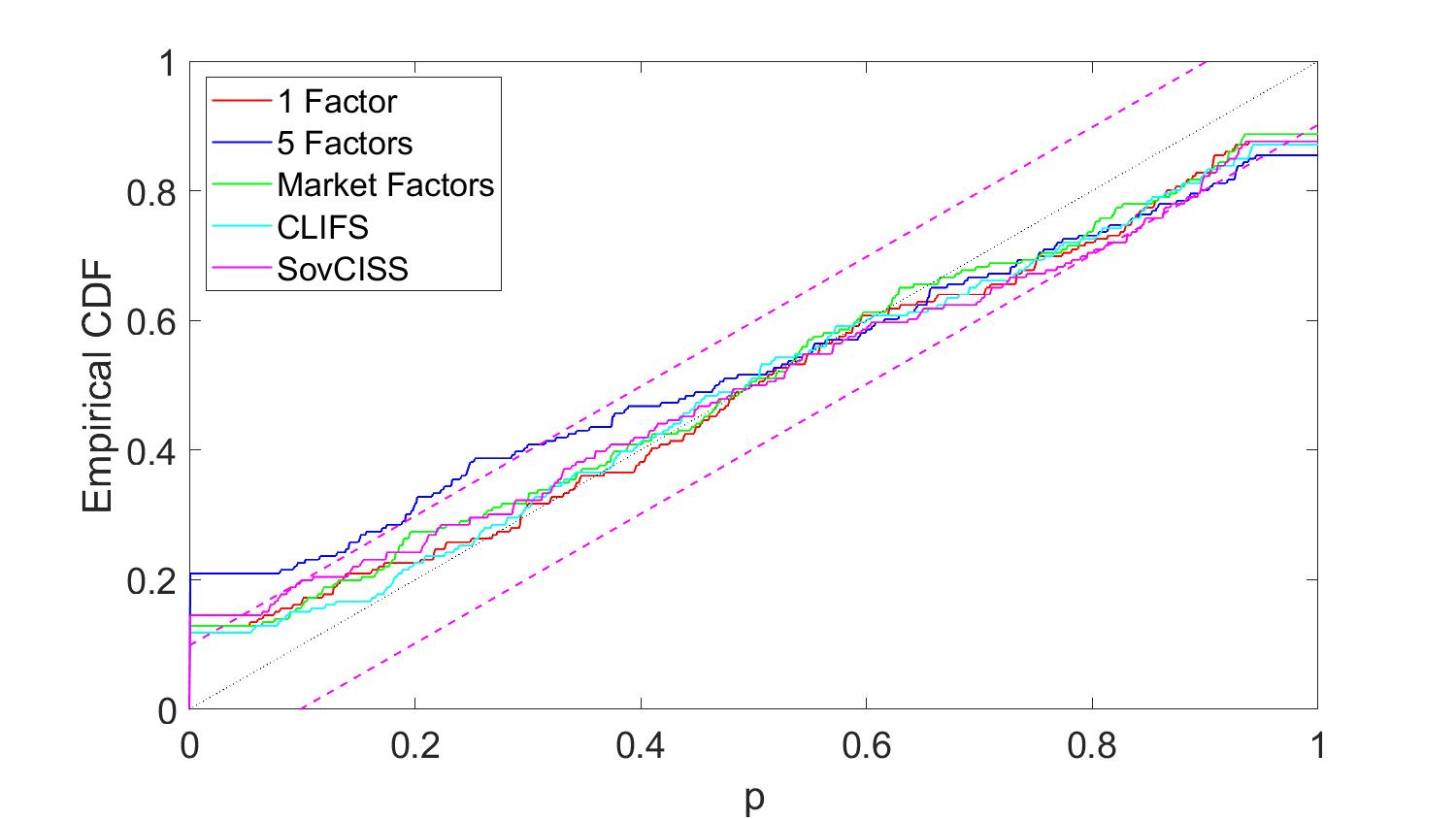}
    \caption{PITS for h=1}
    \label{fig:RAWPITS_h1}
    \end{subfigure}
\hfill
    \begin{subfigure}[b]{0.49\textwidth}
        \includegraphics[width=\textwidth]{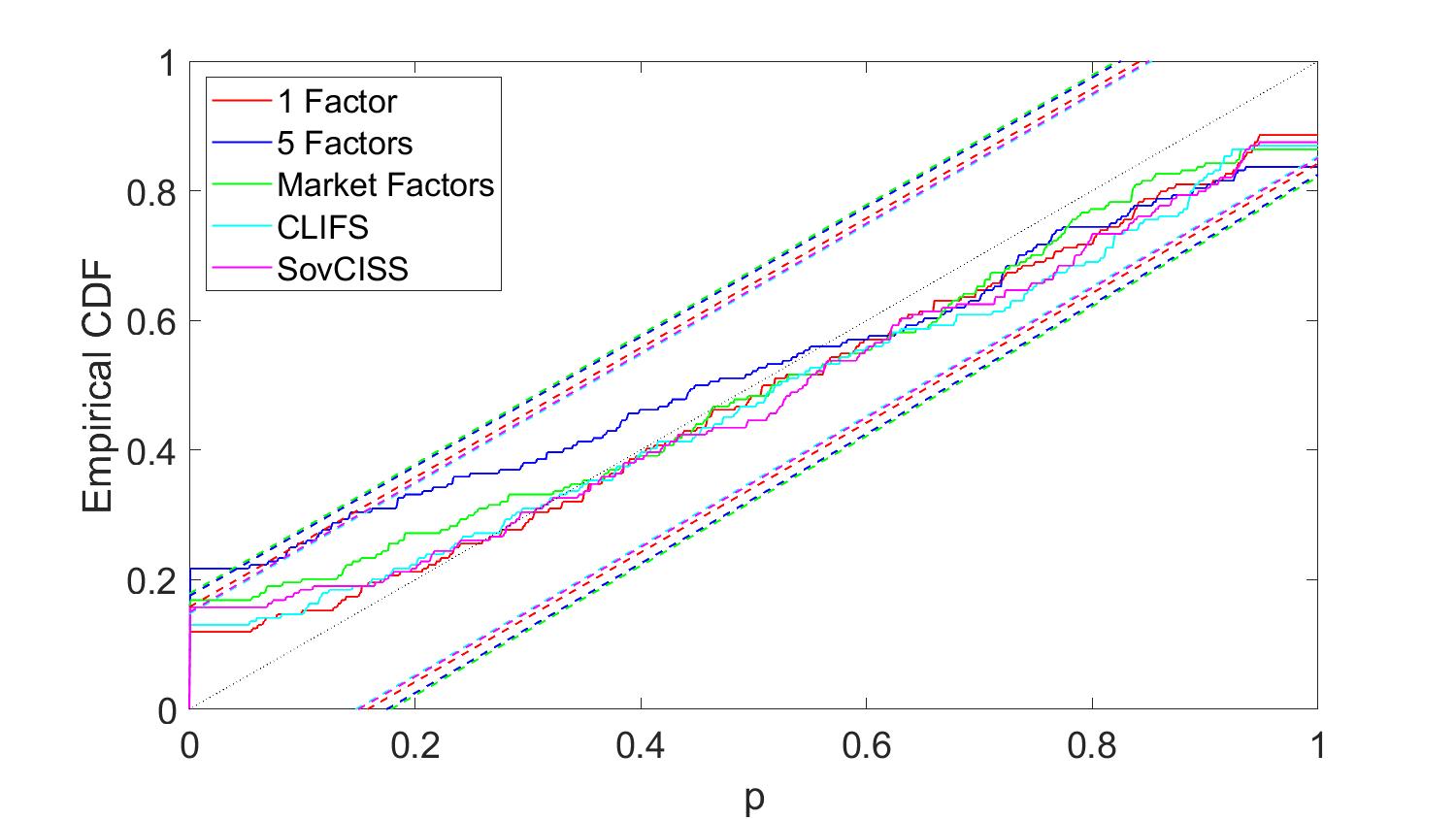}
    \caption{PITS for h=3}
    \label{fig:RAWPITS_h3}
    \end{subfigure}
\hfill
    \begin{subfigure}[b]{0.49\textwidth}
        \includegraphics[width=\textwidth]{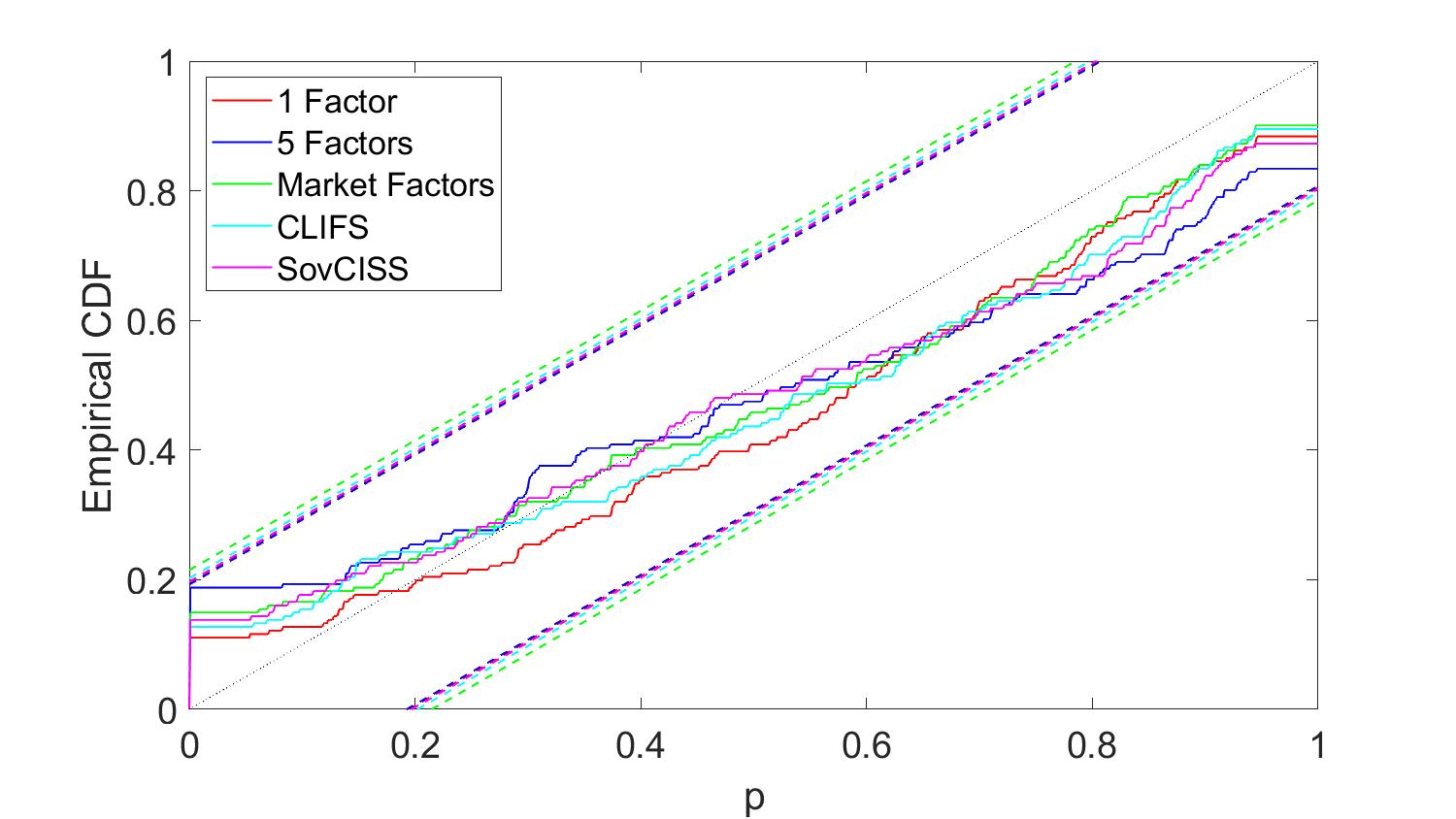}
    \caption{PITS for h=6}
    \label{fig:RAWPITS_h6}
    \end{subfigure}
    \hfill
    \begin{subfigure}[b]{0.49\textwidth}
        \includegraphics[width=\textwidth]{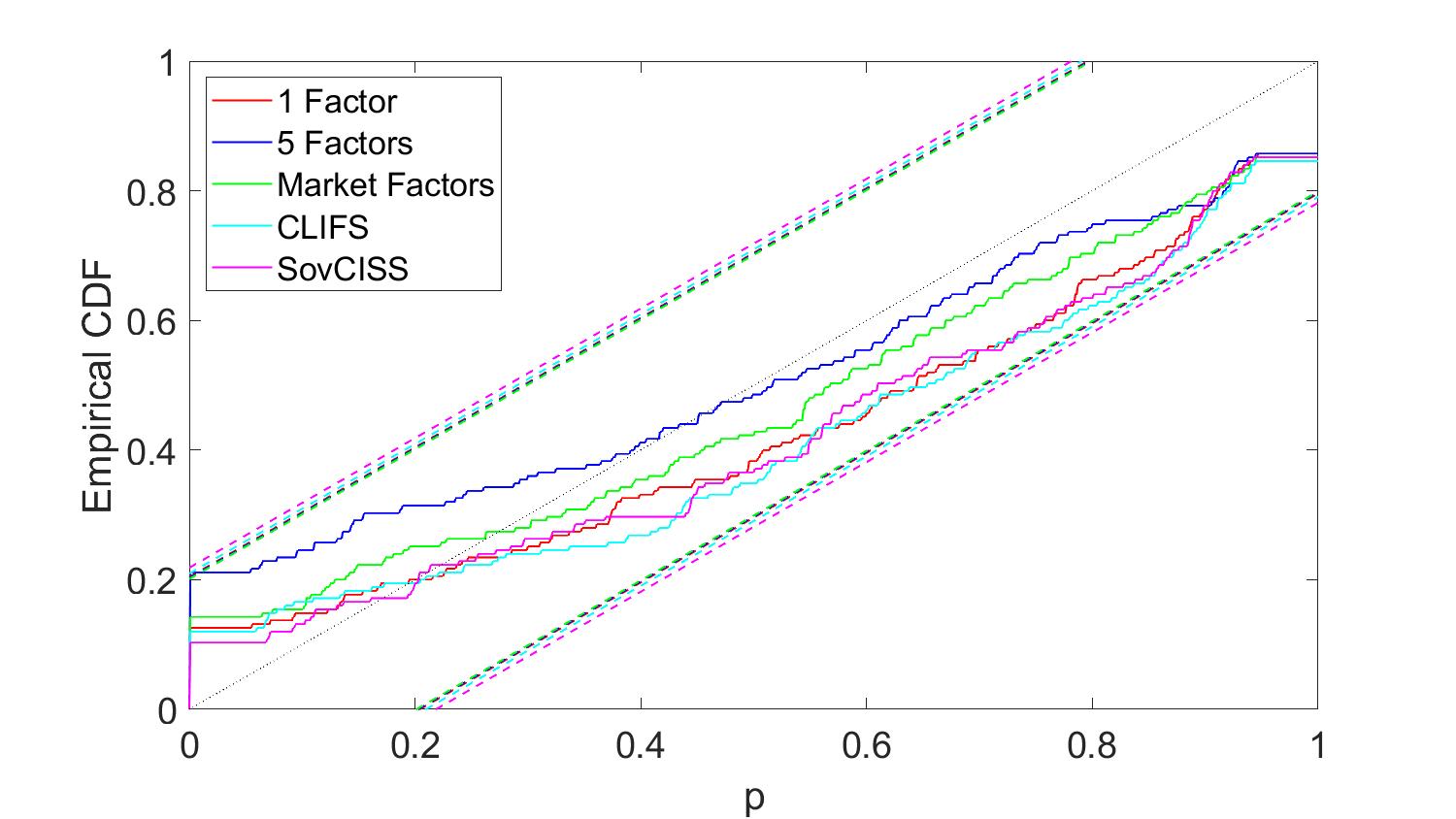}
    \caption{PITS for h=12}
    \label{fig:RAWPITS_h12}
    \end{subfigure}
    \caption{PITS for the different models}
    \label{fig:RAWPITS}
\end{figure}

To evaluate whether the forecasted densities are well calibrated, we present the Probability Integral Transform (PIT) of the forecasted quantiles in figure (\ref{fig:RAWPITS}). We also show the bands of \citet{rossi2019alternative} for the different PITS. The figures reveal that the forecasted densities are well calibrated except in the case for the left tail for h=1 for the 5 factor model. As such the figure provides an explanation of why the 5 factor model yields inferior forecast performance compared to the 1 factor model: the left tail of the 1 factor model is better calibrated than the 5 factor model's.

Given the number of factors are different for the different models, one might argue that better performance is on account of additional covariates in the model. While the AIC and BIC penalise the number of covariates in the fit evaluation, the out-of-sample measures do not. To this end we will also combine the factors, so that only one covariate is included in the growth-at-risk exercise.

Figure (\ref{fig:combfact}) shows the combined statistical factors, and the combined market factors. For completeness on this figure we also show the model with only 1 factor estimated. The figure shows the the different combined factors have large peaks at roughly the same time. Nevertheless, the combined 5 factor model tends to increase earlier than the combined market factor and the single factor model. This can be useful as it showcases the potential for the statistical factor model to act as an early warning system for the policy maker. Furthermore, just like before, the combined statistical factor is less prone to sudden `spikes' from one period to the next. This is another attractive feature of the statistical factors, as it is less likely to lead to situations of identifying stress events falsely.

\begin{table}[]
\centering
\caption{In- and out-of-sample performance of the GaR with different combined risk measures}
\label{tab:Combresult}
\begin{tabular}{llccccc}
\hline
 &  & 1 Factor & 5 Factors & Market Factors & CLIFS & SovCISS \\
 &  &  & (Combined) & (Combined) &  &  \\ \hline
h=1 &  &  &  &  &  &  \\
 & AIC & \textbf{76.001} & 76.827 & 76.214 & 76.857 & 76.647 \\
 & BIC & \textbf{76.280} & 77.105 & 76.493 & 77.136 & 76.926 \\
 & $w_{centre}$ & \textbf{0.054} & 0.055 & 0.055 & 0.055 & 0.055 \\
 & $w_{left}$ & \textbf{0.085} & 0.089 & 0.086 & 0.089 & 0.088 \\
 %& $w_{right}$ & \textbf{0.139} & 0.140 & 0.141 & 0.141 & 0.141 \\ 
 \hline
\multicolumn{2}{l}{h=3} &  &  &  &  &  \\
 & AIC & \textbf{92.307} & 92.839 & 92.343 & 93.481 & 92.841 \\
 & BIC & \textbf{92.587} & 93.120 & 92.624 & 93.762 & 93.121 \\
 & $w_{centre}$ & 0.132 & \textbf{0.131} & 0.133 & 0.136 & 0.132 \\
 & $w_{left}$ & \textbf{0.210} & 0.223 & 0.218 & 0.230 & 0.222 \\
 %& $w_{right}$ & 0.333 & \textbf{0.328} & 0.335 & 0.339 & 0.333 \\ 
 \hline
\multicolumn{2}{l}{h=6} &  &  &  &  &  \\
 & AIC & 99.394 & 98.658 & 98.669 & 99.854 & \textbf{98.253} \\
 & BIC & 99.677 & 98.942 & 98.952 & 100.137 & \textbf{98.536} \\
 & $w_{centre}$ & 0.189 & \textbf{0.186} & 0.188 & 0.198 & \textbf{0.186} \\
 & $w_{left}$ & 0.324 & 0.324 & 0.324 & 0.340 & \textbf{0.310} \\
 %& $w_{right}$ & 0.466 & 0.461 & \textbf{0.460} & 0.484 & 0.464 \\ 
 \hline
 \multicolumn{2}{l}{h=12} &  &  &  &  &  \\
 & AIC & 101.900 & 100.142 & 101.774 & 101.254 & \textbf{97.207} \\
 & BIC & 102.188 & 100.430 & 102.062 & 101.542 & \textbf{97.496} \\
 & $w_{centre}$ & 0.237 & 0.213 & 0.239 & 0.231 & \textbf{0.184} \\
 & $w_{left}$ & 0.399 & 0.380 & 0.401 & 0.385 & \textbf{0.306} \\
 %& $w_{right}$ & 0.577 & 0.513 & 0.588 & 0.563 & \textbf{0.447} \\ 
 \hline
\end{tabular}
\end{table}

\begin{figure}
    \centering
    \begin{subfigure}[b]{0.49\textwidth}
        \includegraphics[width=\textwidth]{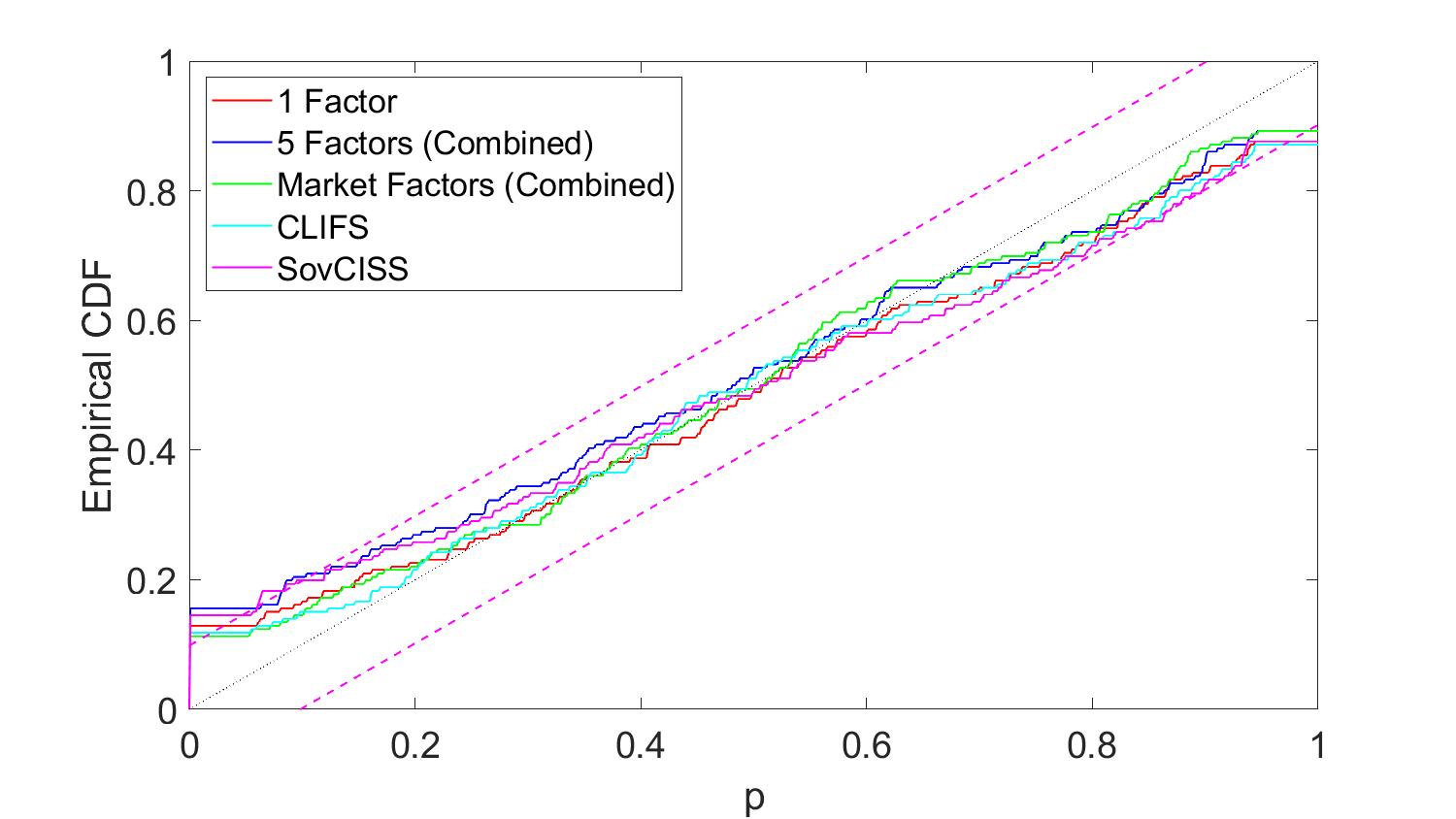}
    \caption{PITS for h=1}
    \label{fig:COMBPITS_h1}
    \end{subfigure}
\hfill
    \begin{subfigure}[b]{0.49\textwidth}
        \includegraphics[width=\textwidth]{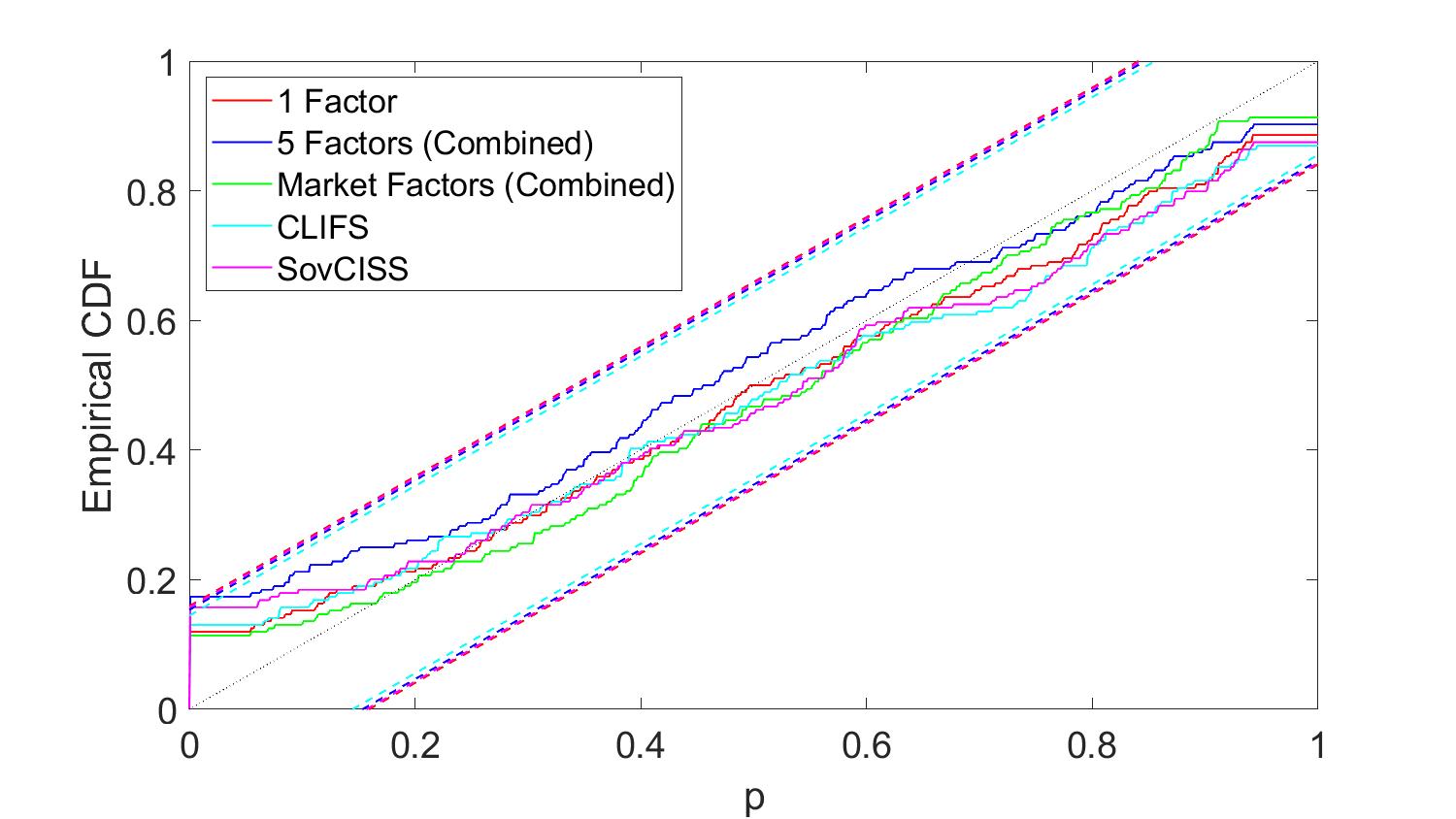}
    \caption{PITS for h=3}
    \label{fig:COMBPITS_h3}
    \end{subfigure}
\hfill
    \begin{subfigure}[b]{0.49\textwidth}
        \includegraphics[width=\textwidth]{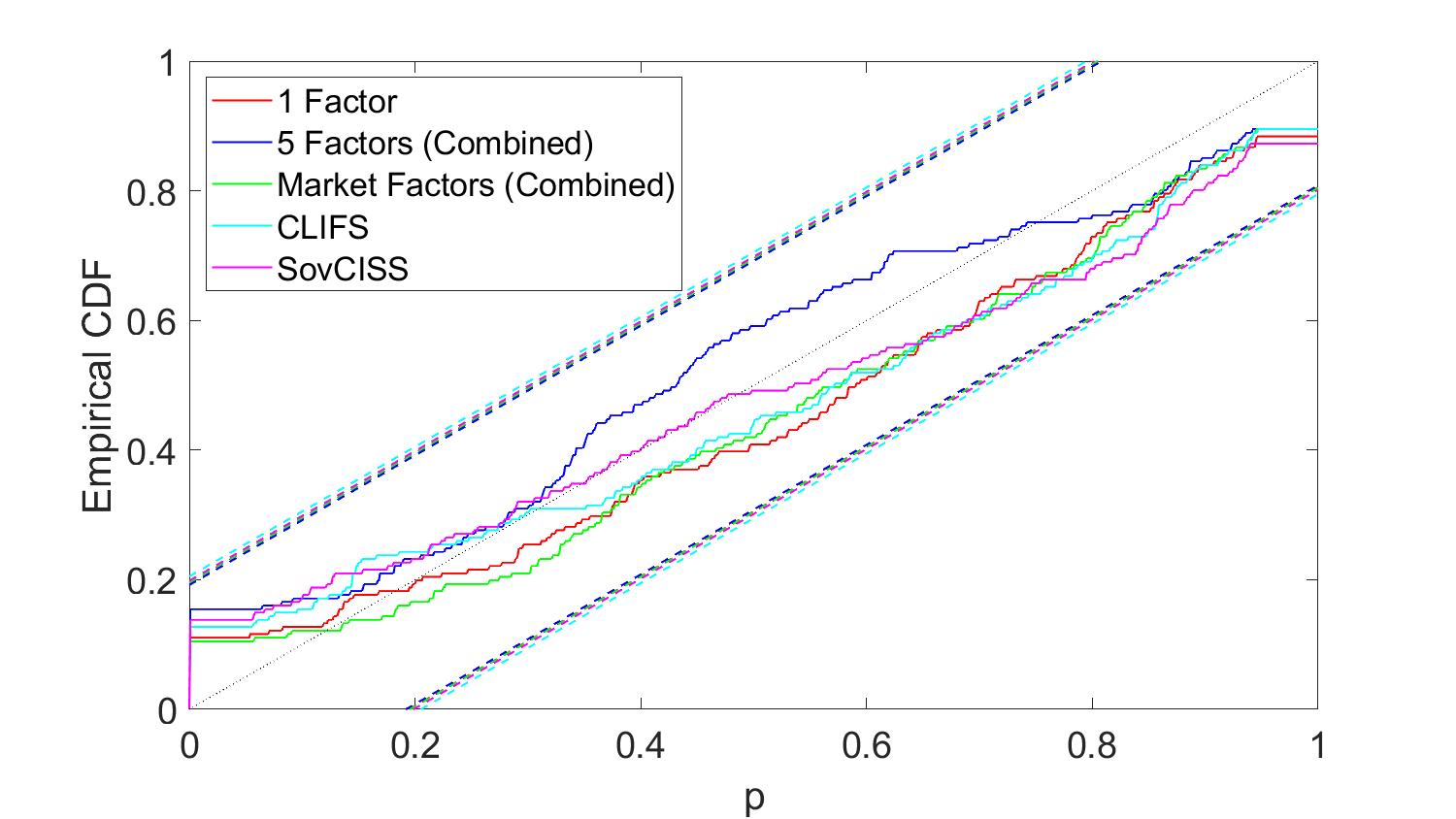}
    \caption{PITS for h=6}
    \label{fig:COMBPITS_h6}
    \end{subfigure}
    \hfill
    \begin{subfigure}[b]{0.49\textwidth}
        \includegraphics[width=\textwidth]{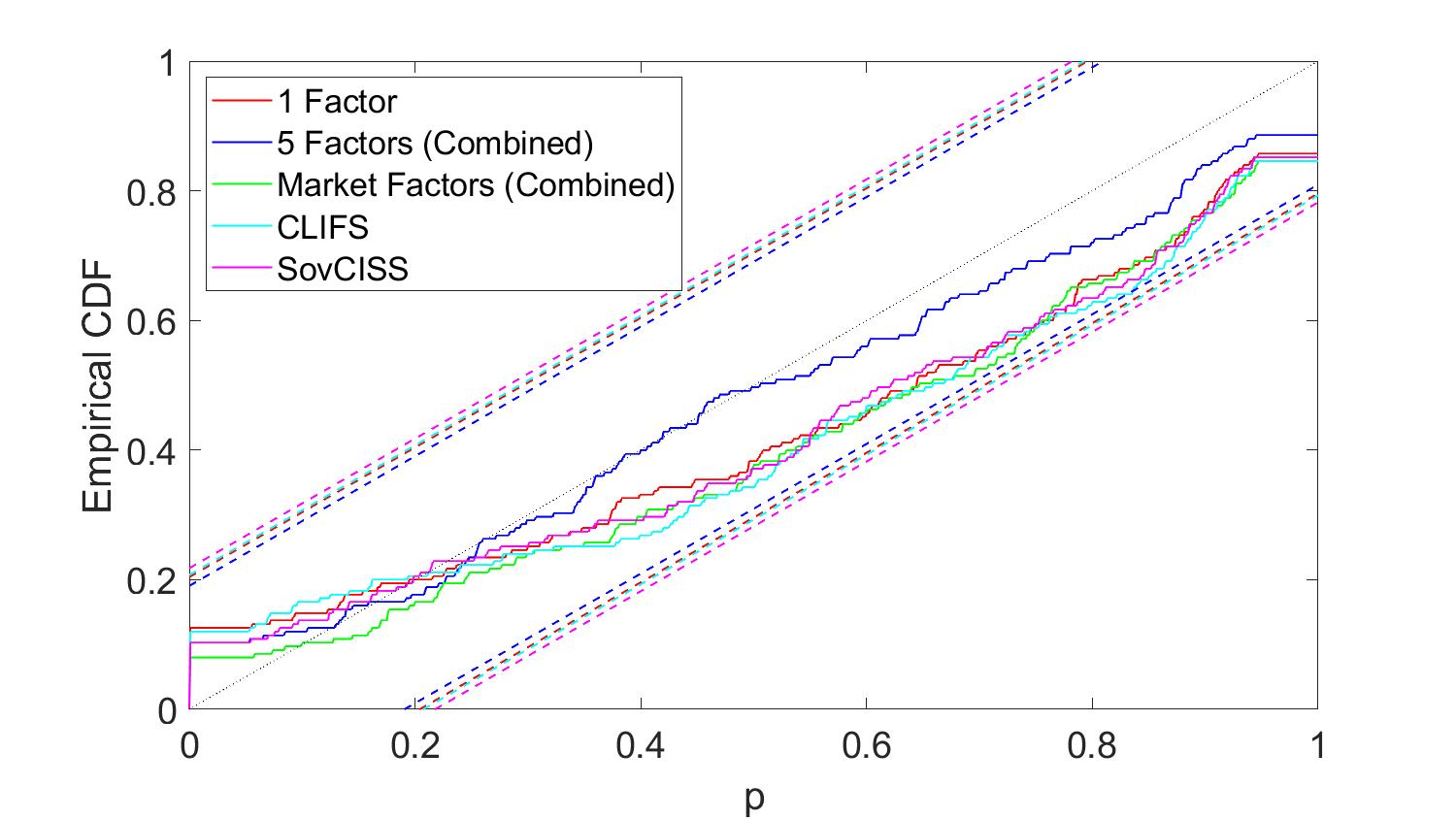}
    \caption{PITS for h=12}
    \label{fig:COMBPITS_h12}
    \end{subfigure}
    \caption{PITS for the different models (Combined factors)}
    \label{fig:COMBPITS}
\end{figure}

Table (\ref{tab:Combresult}) shows the results of the different combined factors models. The table reveals that the models with the combined 5 factor model and the combined market factor model perform worse than before. As such, aggregation does have a noticable impact on model performance. Nevertheless, the overall message that factor models perform better at shorter forecast horizons remains intact, with the SovCISS only being the best model for half a year ahead forecast onwards. Furthermore, just like before, the CLIFS yields the worst performance especially at shorter horizons. When it comes to model calibration, all the combined factors models are well calibrated at all forecast horizons, as shown in figure (\ref{fig:COMBPITS}).

In summary, we show that the factor methodology is a potent way to create measures that describe financial stress. In particular, we show that it is optimal to pool all the chosen variables together and let the method select the optimal number of factors given the test described in \citet{pena2006nonstationary}. Doing so allows one to identify factors that jointly describe the financial market. With a growth-at-risk exercise we show that these factors are particularly potent when it comes to short-run forecasts of the left tail. Nevertheless, we also show that our proposed method fairs worse than the SovCISS at longer forecast horizons, which highlights that different ways of modelling financial stress uncover different aspect of stress. To this end we propose that the proposed factor based financial stress measure is used along with pre-existing stress measures. We also highlight that aggregating the factors yields worse performance for the model. We note however, that in this paper we used a simple arithmetic average to obtain aggregated factors, and we leave it for future research to identify better ways to aggregate the financial stress factors.

\newpage

\section{Conclusion}
We extend the findings of previous research, particularly the conclusion drawn by FISS \citep{szendrei2020fiss}, which suggests that non-stationary factors offer more effective financial stress indices. Building upon this, we use a robust mathematical approach introduced by \citet{pena2004forecasting} and \citet{pena2006nonstationary} to determine the optimal number of factors required in our model. By doing so, we aim to enhance the accuracy and reliability of our stress indices.  Specifically, we focus on constructing a factor-based stress index tailored for the UK financial system.

One notable advantage of our statistical factors is their greater persistence compared to individual variables. This results in smoother factors, which are crucial for avoiding false positives in policy decision-making processes. By prioritising factors with higher persistence, we aim to provide policymakers with more reliable indicators of financial stress, thereby contributing to more informed policy responses.

In evaluating the performance of our stress indices, we employ a growth-at-risk metric. This evaluation method allows us to assess the effectiveness of our indices in capturing downside risk stemming from vulnerabilities in the financial system. By utilising growth-at-risk, we ensure a comprehensive assessment of all stress indices' performance, enabling us to identify key differences between the various stress indices.

Our research reveals an interesting finding regarding the effectiveness of different stress indices at varying time horizons. While factor-based indices demonstrate efficacy in short-term forecasting, the Composite Indicator of Systemic Stress (CISS) proves highly effective for longer-term forecasting, particularly at a one-year horizon. This disparity underscores the importance of considering different methodologies in stress index construction, as each method reveals distinct aspects of risk. Specifically, the CISS methodology excels in identifying systemic risk, where multiple markets are collectively affected, leading to prolonged crises. Consequently, we advocate for viewing the various stress index methodologies not as substitutes but as complementary tools in understanding the multifaceted nature of financial stress. This underscores the complexity of measuring financial stress and highlights the need for a nuanced, multifaceted approach to risk assessment in financial markets.

%\tibi{Key contributions: 
%\begin{enumerate}
%    \item We build on the finding of FISS, that non-stationary factors provide better financial stress indices. We now use a mathematically robust way to select the number of factors needed in our model!
%    \item Our statistical factors are more persistent than a lot of the individual variables: this leads to smoother factors. This is useful as false positives are not good for policy making.
%    \item We use growth-at-risk as evaluation for the performance of the stress indices
%    \item Finding that different stress indices effective at different horizons: Factor based indices work great for short horizon, but CISS works great for 1-year ahead horizon. This highlights that although the data these measures are constructed from are similar, the method of aggregation reveals different aspects of risk. In particular the methodology of the CISS helps reveal cases of systemic risk, i.e. situations where different markets are jointly impeded. Such crises have a longer lasting impact which explains power to forecast downside risk at longer horizons. As such, the different stress index methodologies are not substitutes but should instead be looked as complements. This also highlights that it is difficult to have an overarching measure of financial stress, stemming from the fact that different types of financial stress exists.
%\end{enumerate}}

\pagebreak

%%TC:ignore
\bibliographystyle{chicago}
%\addcontentsline{toc}{chapter}{Bibliography}
%\pagestyle{ref}
%\bibliography{reference.bib}
\bibliography{main.bbl}
%%TC:endignore

\pagebreak

%\section{Graphs}
%\include{Chapters/appfiguresincl}

\appendix 
\section{Appendix}
\subsection{Technical Appendix}

Depending on the assumptions on the model the following limit theorems hold \citep{pena2006nonstationary}. Let's suppose that the non-stationary factor model is true:

\begin{equation*}
    \begin{split}
        X_t=Lf_t+\varepsilon_t\\
        \Phi(b)f_t=d+\Theta(B)u_t\\
        (1-B)^d f_{1,t}=c+v_t \\
        v_t=\Psi(B)u_{1,t}
    \end{split}
\end{equation*}

\noindent and $c=0$, $C_X(k)$ is the generalised covariance matrix, for $k=1,2,...,K$ and K is small compared to the sample size, that is if $T\rightarrow \infty$, $K/T\rightarrow 0$. Then

\begin{enumerate}
    \item The generalised sample covariance matrices, $C_X(k)$ converge weakly to a random matrix $\Gamma_X$, for $k=1,2,...,K$ where limits are taken as $T\rightarrow \infty$, and $\Gamma_X$ is defined as:
    \begin{equation*}
         \Gamma_X=L_1\Psi(1)\Sigma_1^{1/2}\Big(\int^1_0 V_{d-1}(\tau)V_{d-1}(\tau)'d\tau \Big) \Sigma_1^{1/2'}\Psi(1)'L_1'
    \end{equation*}
    where $V_d(\tau)=F_d(\tau)-\int^1_0F_d(\tau)d\tau$ is the d times integrated Brownian motion, and it is defined recursively by $F_d(\tau)=\int^\tau_0F_{d-1}(s)ds$, $d=1,2,...$ with $F_0(\tau)=W(\tau)$, the $r_1$-dimensional standard Brownian motion
    \item $\Gamma_X$ has $r_1$ eigenvalues greater than zero almost surely and $m-r_1$ equal to zero.
    \item The eigenvectors corresponding to the $r_1$ eigenvalues of $\Gamma_X$ greater than zero are a basis of the space spanned by the columns of the loading sub-matrix $L_1$.
\end{enumerate}

The following statement describes the convergence results when the common factors have drifts.

For the non-stationary factor model given above with $c\neq0$ and defining $C_X(k)$ as the generalised covariance matrix with $D=1$, for $k=1,2,...,K$ where $K/T\rightarrow 0$ and limits are taken as $T\rightarrow\infty$:
\begin{equation*}
    C_X(k)\overset{P}{\rightarrow}qL_1cc'L_1'
\end{equation*}

where q is a constant depending on d. It is obvious, that in this case the limit is non-stochastic and is driven by the drift term.

After clarifying the limiting behaviour of the generalised covariance, we apply the continuous mapping theorem of \citet{andersson1983distribution} to underpin that the eigenvalues and eigenvectors of the generalised empirical covariance matrix are maximum likelihood estimates of those of the limiting covariance matrices.

The number of common non-stationary factors could be estimated as the number of eigenvalues of $C_X (k)$ converging weakly to the $r_1$ nonzero eigenvalues of their limit matrix $\Gamma_X$. Since $C_X(k)\overset{d}{\rightarrow}\Gamma_X$ and the eigenvalues are continuous functions of the covariance matrix, we can apply the continuous mapping theorem to prove that the ordered eigenvalues of $C_X (k)$ converge weakly to those of $\Gamma_X$. The standard chi-square test will be used to estimate the number of common factors whether they are stationary or not.

\subsection{Tables}

\begin{table}[h!]
\centering
\caption{Variables used for the monthly GDP model}
\label{tab:varibGDP}
%\resizebox{\textwidth}{!}{%
\begin{tabular}{l|l|l}
\multicolumn{1}{c|}{\textbf{Variable}} & \multicolumn{1}{c|}{\textbf{Frequency}} & \multicolumn{1}{c}{\textbf{Transformation}} \\ \hline
Real gross domestic product (production) & Quarterly & Annualised growth rate \\
Real gross domestic product (expenditure) & Quarterly & Annualised growth rate \\
Industrial production index (manufacturing) & Monthly & Growth rate \\
FTSE 100 & Monthly & Growth rate \\
Civilian unemployment rate & Monthly & Growth rate \\
CPI: All items & Monthly & Growth rate \\
Retail sales & Monthly & Growth rate \\
10-year government bond & Monthly & Level \\
3-month government bond & Monthly & Level \\
Base rate & Monthly & Level
\end{tabular}%
%}
\end{table}

\end{document}